\documentclass{jfm}
\usepackage{graphicx}
\usepackage{mathtools}
\usepackage{epstopdf, epsfig}
\usepackage{subcaption,color}
\definecolor{cincinnati-red}{RGB}{190,0,0}
 
\def\rev #1{{\textcolor{black}{#1}}}   
\DeclareMathOperator{\tr}{tr}

\shorttitle{Shear-induced sedimentation of a sphere  in  a yield stress fluid}
\shortauthor{M. Sarabian, M.E. Rosti, L. Brandt and S. Hormozi}

\title{\rev{Numerical simulations of a sphere settling in simple shear flows of yield stress fluids}}

\author{Mohammad Sarabian\aff{1},
  Marco E. Rosti\aff{2,3},
  Luca Brandt\aff{2}
 \and Sarah Hormozi\aff{1}
\corresp{\email{hormozi@ohio.edu}}}

\affiliation{\aff{1}Department of  Mechanical Engineering, Ohio University, 251 Stocker Center, Athens, Ohio, USA
\aff{2}Linn$\acute{\rm e}$ Flow Centre and SeRC (Swedish e-Science Research Centre), KTH Mechanics, SE 100 44 Stockholm, Sweden
\aff{3} Complex Fluids and Flows Unit, Okinawa Institute of Science and Technology Graduate University, 1919-1 Tancha, Onnason, Okinawa 904-0495, Japan} 
\begin{document}

\maketitle

\begin{abstract}
We perform $3$D numerical simulations to investigate the sedimentation of a single sphere in the absence and presence of a simple cross shear flow in a yield stress fluid with weak inertia. In our simulations, the settling flow is considered to be the primary flow, whereas the linear cross shear flow is a secondary flow with amplitude $10\%$ of the primary flow. To study the effects of elasticity and plasticity of the carrying fluid on the sphere drag as well as the flow dynamics, the fluid is modeled using the elastovisco-plastic (EVP) constitutive laws proposed by \cite{saramito2009new}. The extra non-Newtonian stress tensor is fully coupled with the flow equation and the solid particle is represented by an immersed boundary (IB) method. Our results show that the fore-aft asymmetry in the velocity is less pronounced and the negative wake disappears when a linear cross shear flow is applied. We find that the drag on a sphere settling in a sheared yield stress fluid is reduced significantly as compared to an otherwise quiescent fluid. More importantly, the sphere drag in the presence of a secondary cross shear flow cannot be derived from the pure sedimentation drag law owing to the non-linear coupling between the simple shear flow and the uniform flow. Finally, we show that the drag on the sphere settling in a sheared yield-stress fluid is reduced at higher material elasticity mainly due to the form and viscous drag reduction.  
\end{abstract}


\section{Introduction}

Suspensions of dense particles in a yield stress fluid are ubiquitous in many engineering processes (crude oil, foodstuff transport, cosmetics, microfluidics, mineral slurries, cement pastes, fermentation processes, $3D$ printing, drilling muds, \textit{etc}), natural phenomena (debris flows, lava flows, natural muds, \textit{etc}), and biological systems (physiology, biolocomotion, tissue engineering, \textit{etc}). Yield stress fluids exhibit a yield stress, beyond which they deform as a non-Newtonian viscous liquid while they behave as solids for lower stress levels.  In mentioned practical applications, one is usually dealing with the transport of the suspending particles in yield stress fluids. Hence, in these systems, particle sedimentation occurs in the presence of a shear flow. In Stokes flow of Newtonian fluids, the governing equations of motion are linear thus the settling rate of a spherical particle is not affected by a shear flow superimposed on the settling flow. However, in yield stress fluids, the sphere drag is affected by a cross shear flow as the linearity of the Stokes equations breaks due to the non-linear rheology of the yield stress fluids. The objective of this paper is to investigate the flow  dynamics and drag laws of a sphere settling in yield stress fluids when a cross shear flow exists.
\par 
Sole plastic properties of yield stress fluids can be addressed via adopting ideal yield stress constitutive laws such as Bingham and Herschel-Bulkley models (also known as visco-plastic models). However, experimental works show that elasticity plays an important role in problems involving inclusions in practical yield stress fluids  \cite[]{holenberg2012particle,firouznia2018interaction}. Therefore,  the goal of the present study is twofold: we \rev{do not} only study the nonlinear coupling of a settling and shear flow around a sphere but also take into account the plastic  as well as  the elastic properties characterising many practical yield stress fluids. Hence, in this introduction, we first give a review of several  works that have been undertaken on flows around a 3D or a 2D particle immersed in a visco-plastic fluid, see also \cite{maleki2015macro}. Second,  we review the works addressing the role of elasticity  in a yield stress fluid and a pure sedimentation problem. Third, we review a limited recent number of works that investigate shear induced sedimentation of particles in viscoelastic fluids.

\par
The problem of a single spherical particle settling in an ideal  yield stress fluid has been extensively studied theoretically and numerically \rev{\cite[see e.g.][]{andres1960equilibrium,yoshioka1971creeping, beris1985creeping,blackery1997creeping,beaulne1997creeping,liu2002convergence,yu2007fictitious}}. Two main numerical difficulties arise in solving this problem; handling a freely-moving particle, and the numerical treatment of the yield stress constitutive equations as the effective viscosity becomes infinite at the yield surface and within a solid region. 

\par

To eliminate the first difficulty, this problem has primarily been studied by fixing the particle and imposing an external flow while the confining walls are translating with the same velocity of the medium. This is called the \textit{resistance} problem (i,e, the flow past a fixed obstacle). Most of these simulations were performed in 2D, either by assuming the obstacle to be an infinitely long circular cylinder \rev{\cite[see e.g.][]{zisis2002viscoplastic,roquet2003adaptive,de2003viscoplastic,mitsoulis2004creeping,tokpavi2008very,putz2010creeping,wachs2016particle,chaparian2017yield,chaparian2017cloaking,ouattara2018drag}} or a spherical particle by imposing 2D axisymmetric boundary conditions  \rev{\cite[e.g.][]{beris1985creeping,blackery1997creeping,beaulne1997creeping,liu2002convergence}.} To our knowledge, only \rev{two} studies exist that solves the problem of a particle settling in visco-plastic fluids \cite[][]{yu2007fictitious} \rev{and creeping flow of Bingham plastics around translating objects \cite[][]{sverdrup2019embedded} using fully 3D numerical simulations. Recently, the flow past a rotating sphere in a Bingham plastic fluid has been investigated for a wide range of material property and Reynolds number by \cite{pantokratoras2018flow}.}

\par  
Two different approaches are implemented to remove the second difficulty; the regularization method and the Augmented-Lagrangian method. The former removes the inherent discontinuity of an ideal yield stress constitutive equation \cite[see e.g.][]{bingham1922fluidity,herschel1926measurement} by approximating the solid region as a fluid with an extremely high viscosity. \cite{frigaard2005usage} provides a review on different regularization schemes. The Augmented-Lagrangian scheme, on the other hand, maintains the true constitutive equation but requires the solution of an often expensive minimisation problem. For more details on this method the reader is referred to \cite{glowinski2011numerical}. 
\par 
\rev{When a sphere is falling in a yield stress fluid, the stresses decay as we move away from the surface of the particle and they may become smaller than the yield stress. Therefore,  there exists a rigid envelope (a plug or an unyielded region) surrounding the liquid (yielded) zone around the particle \cite[][]{volarovich1953theory, ansley1967motion}. Moreover,   two unyielded triangular polar caps form at the front and rear stagnation points of the sphere \cite[][]{beris1985creeping, blackery1997creeping,beaulne1997creeping}. It is noteworthy to mention that in the literature different terms have been used to refer to the location of the polar caps such as the leading and trailing parts of the spheres or north and south poles of the sphere.  For the 2D cases  (i.e.,  cylinders in an infinite domain) it has been shown that,  in addition to the rigid envelope and the stagnant regions attached to the front and rear of the cylinder, two counter rotating solid islands may form at both sides of the cylinder's equator in the fluid zone \cite[see e.g.][]{de2003viscoplastic, chaparian2017yield}. \cite{ansley1967motion} postulated the shape of the plug regions around a falling sphere and pointed to plasticity theory as a tool for tackling problems involving yield stress fluids. However, only more recently researches have made systematic use of plasticity theory and especially the slipline method for analyzing the yield limit in 2D problems involving yield stress fluids \cite[][]{liu2016two,chaparian2017cloaking,chaparian2017yield,  balmforth2017viscoplastic}.}

\par
\rev{Wall effects on the evolution of the yielded/unyielded zones has been investigated numerically and experimentally.  These studies include either  a particle settling in a tube filled with a visco-plastic fluid  \cite[]{blackery1997creeping,beaulne1997creeping} or the flow of a visco-plastic fluid past a 2D cylinder (a  circular long cylinder) located  between two parallel plates \cite[see e.g.][]{zisis2002viscoplastic,mitsoulis2004creeping,ouattara2018drag}. Generally, it is found that at a fixed confinement ratio, the yielded zone surrounding the particle  shrinks with increasing yield stress leaving thin viscous  layers around the particle. A well-resolved series of simulations for the case of a settling circular disk shows that the thin viscous layers resemble a cross-eyed owl \citep{wachs2016particle}. At  fixed yield stress, increasing the confinement ratio results in the extension of the yielded region around the particle and,  eventually, its interaction with the walls.}
\par 
\rev{The drag exerted on a particle settling in an ideal yield stress fluid is found to be a function of both the Bingham number and the confinement ratio. The sphere drag is an increasing function of Bingham number ($Bi$) at fixed confinement ratio \cite[see e.g.][]{atapattu1995creeping,blackery1997creeping,tabuteau2007drag,ahonguio2014influence}. \cite{blackery1997creeping} reported the Stokes drag coefficient (i.e., the ratio of the total drag force to the Stokes drag on a sphere settling in a Newtonian fluid) as a function of the Bingham number (i.e., the ratio of  the yield stress to viscous stresses) and showed an enhancement of the Stokes drag coefficient with increasing the confinement ratio. For large Bingham numbers $(Bi\ge 10)$, however,  the Stokes drag coefficient is independent of the confinement ratio. At large enough $Bi$ numbers, the envelope of the yielded zone around a particle is encapsulated by an outer plug region attached to the boundaries. Thus, at this point, a further reduction of the confinement does not affect the flow dynamics and drag forces. These results are qualitatively valid  for  cases of visco-plastic fluids flowing around 2D cylinder in a duct  \cite[]{zisis2002viscoplastic,mitsoulis2004creeping}. Wall slip also alters the flow dynamics and consequently the drag forces in flows of yield stress fluids over solid boundaries.  In  experimenting with practical visco-plastic materials, the slip at the wall is unavoidable \cite[see e.g.][]{meeker2004slip,holenberg2012particle}.  \cite{de2003viscoplastic} showed that the wall slip reduces  the drag force exerted by a visco-plastic fluid on a 2D cylinder in an infinite medium.}
\par 
\rev{It is of practical interest to know the force required to move a body immersed in a yield stress fluid. We refer the reader to the early and historical work of \cite{boardman1960yield}. In sedimentation flows,  a particle does not fall when the yield stress resistance is larger than the buoyancy stress. A dimensionless number called  gravity number $Y_{G}$ is defined as the ratio between material yield stress and the buoyancy stress. There exists a critical value of  $Y_{G}$  beyond which the particle ceases to move and it is entrapped within the yield stress fluid. The critical value of  $Y_{G}$ depends on the geometry of the immersed object \citep{ovarlez2019lectures} and it  has been found numerically for a 2D circle \citep{tokpavi2008very}, a sphere \citep{beris1985creeping}, a 2D square \citep{nirmalkar2012creeping} and a 2D ellipse \citep{putz2008settling}. Experimental attempts also have been made to determine the critical value of $Y_{G}$  for objects with different shapes  \cite[see e.g.][]{jossic2001drag,tabuteau2007drag,tokpavi2009experimental,ahonguio2015motion}. There exists a quantitative disagreement between the numerical results and experimental results due to the unavoidable wall-slip effects in experiments as well as owing to the discrepancy between the rheological behavior of the practical yield stress fluids and the ideal yield stress laws used in the theoretical and computational work.  To this end, one of the objectives of the present work is to conduct a numerical study of the flows of yield stress fluids around a sphere using a more realistic constitutive law, i.e.\ including some fluid elasticity.} 

\par 
In many soft materials, (e.g. emulsions, foams, gels, colloidal pastes, \textit{etc}), elasticity is found to play an important role in the dynamics of the flow  around \rev{a single sphere or  2D cylinder  \cite[see e.g.][]{atapattu1995creeping,gueslin2006flow,dollet2007two,tabuteau2007drag,putz2008settling,holenberg2012particle,ahonguio2014influence,ouattara2018drag}, particles of various shapes \citep{jossic2001drag}, dilute suspensions \citep{hariharaputhiran1998settling}, concentrated suspensions \citep{coussot2002coexistence}, and bubbles rising in yield stress fluids \cite[see e.g.][]{sikorski2009motion,fraggedakis2016velocity}}. Thus, to accurately predict the behavior of practical visco-plastic fluids, one must also take the elasticity effects into account. These materials are called elastoviscoplastic (EVP). There are a number of different constitutive equations proposed in the literature to model EVP fluids \cite[e.g.][]{de2007dimensionless,saramito2007new,benito2008elasto,saramito2009new,dimitriou2013describing,dimitriou2014comprehensive,geri2017thermokinematic}. For more information on the details of these models and their implementation, the reader is referred to \cite{izbassarov2018computational}. Several modelling and numerical work have implemented the constitutive laws of EVP materials to quantitatively capture the flow characteristics of practical yield stress fluids \cite[e.g.][]{cheddadi2011understanding,cheddadi2013new,fraggedakis2016yielding}.

\par 
The problem of particle settling in an EVP fluid confined in a cylindrical container is studied through axisymmetric finite element computations by \cite{fraggedakis2016yielding}. The numerical results show that the loss of fore-aft symmetry in the velocity field around the sphere and the appearance of the negative wake downstream of the sphere settling in a yield stress fluid are due to the elasticity. These phenomena were also observed experimentally \cite[see e.g.][]{holenberg2012particle}. Furthermore, it has been demonstrated that the extent and shape of the yielded/unyielded regions, the particle stoppage criterion, and the sphere drag are influenced by the presence of elasticity in laboratory yield stress fluids \cite[][]{fraggedakis2016yielding}. The  flow field around a single sphere and  trajectories of two spheres  in simple shear flows of practical yield stress fluids (Carbopol gel) were also studies experimentally showing the importance of elastic effects \cite[][]{firouznia2018interaction}. 

\par 
There exists a handful experimental  studies focusing on the shear-induced sedimentation of suspensions of  particles in practical yield stress fluids \cite[see e.g.][]{merkak2009migration,ovarlez2010three,ovarlez2012shear}. \cite{merkak2009migration} studied the shear-induced sedimentation of suspensions of  particles in a pipe flow configuration and proposed the criterion for particle stability in the sheared yield stress fluid. \cite{ovarlez2012shear} examined the settling rate of suspensions of  particles in different yield stress fluids (concentrated emulsions and Carbopol gels). The density mismatched suspensions are sheared in a wide-gap Couette device; the velocity profile and solid volume fraction are measured by employing magnetic resonance imaging  techniques. It is found that, for all the systems, the particles that were stable at rest start sedimenting in the yielded zone as the shear is introduced. Moreover, the experimental results of \cite{ovarlez2012shear} show that the particle settling velocity scaled by the Stokes velocity is an increasing function of the inverse Bingham number. Despite all the experimental results falling into a master curve in the limits of high Bingham number (plastic regime), there exists a discrepancy when $Bi \leq 1$, which might be due to another governing dimensionless number associated with the presence of elasticity in a practical yield stress fluid. 

\par 
A larger number of studies focuses on the effect of shear flow on the particle settling rate in viscoelastic fluids through experimental measurements \cite[see e.g.][]{van1993effects,murch2017growth}, direct numerical simulations \cite[see e.g.][]{padhy2013simulations,padhy2013effect,murch2017growth}, and theoretical analysis \cite[see e.g.][]{brunn1977slow,brunn1977errata,housiadas2012drag,vishnampet2012concentration,housiadas2014rheological,housiadas2014stress,einarsson2017spherical}.  
 \begin{figure}
	\centering
	\includegraphics[trim={0cm 0cm 0cm 0cm},width=0.55\linewidth]{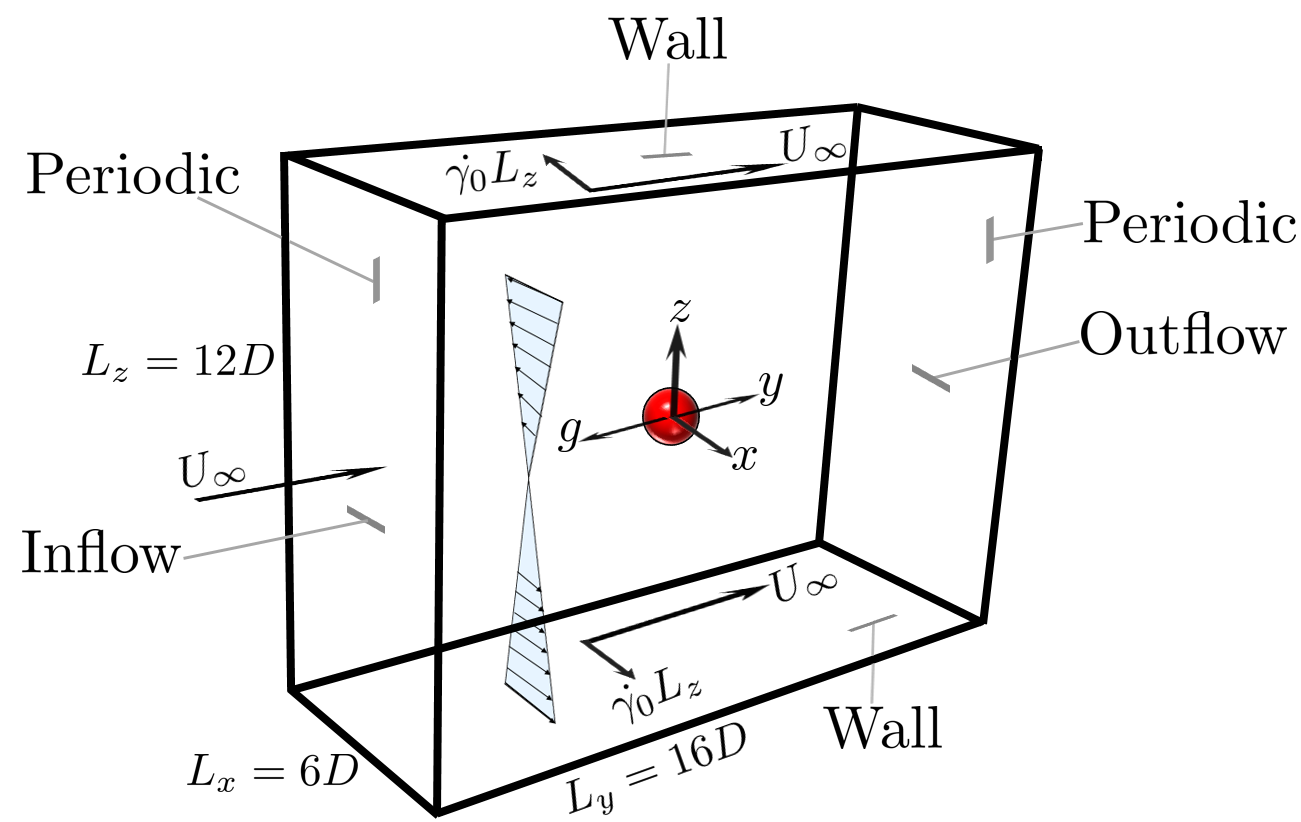} 	
	\caption{Computational domain and boundary conditions}
	\label{fig::fig1}
\end{figure}

\par 
The drag enhancement on the sphere when it is subjected to an externally imposed cross shear flow is verified through direct numerical simulations both in a weakly \cite[][]{padhy2013simulations} and in a highly \cite[][]{murch2017growth} constant-viscosity viscoelastic fluid. In the case of weakly viscoelastic fluids, it has been shown that adding polymers indirectly increases the drag on the sphere through breaking the symmetry in the polymer and viscous stresses on the sphere surface \cite[][]{padhy2013simulations}. Contrary to the weakly viscoelastic fluids, the form drag is a primary cause of the drag enhancement on the particle in the limit of highly viscoelastic fluids when the cross shear flow is coupled with the uniform flow \cite[][]{murch2017growth}. In addition to the elasticity effects, the sphere drag coefficient is also affected by the rheological properties of the viscoelastic fluid such as the degree of shear-thinning and the ratio of polymer to total viscosity \cite[][]{padhy2013effect}.

\par
The main novel contributions of our study can be summarised as follows. First, we validate a new tool for the 3D  numerical simulations of a sphere settling in a quiescent and sheared yield stress fluid at small particle Reynolds number. We formulate the problem using both an ideal yield stress fluid law (Bingham model) and EVP fluid law. The governing equations are solved using our recently developed 3D numerical solver to handle the problems including rigid and deformable particles suspended in complex fluids \cite[]{izbassarov2018computational}. The particle motion in the flow is simulated by means of an efficient Immersed Boundary Method (IBM) coupled with a flow solver for the equations of motions associated with the suspending complex fluids. The mechanical and mathematical modeling, the computational matrix, the boundary conditions, and the numerical scheme of the simulations presented here are reported in section \ref{sec:: problem formulation}. Second, we provide for the first time a 3D and extensive analysis of the velocity, stress  fields and yield surfaces around a sphere settling in yield stress fluids. We consider both quiescent and sheared yield stress fluids and compute the drag force and its individual components, i.e., form drag, viscous drag, polymer drag and inertia drag and discuss their dependency on the elastic and plastic properties of the suspending fluids as well as the  flow dynamics. Third, we show that the flow dynamics and drag forces significantly change when a secondary linear cross shear flow is superimposed. In particular, the nonlinear coupling of the settling and simple shear flow plays a significant role in determining the flow dynamics and consequently the drag force on the sphere, which makes the prediction of sphere drag in a sheared yield-stress fluid difficult (section \ref{sec:: results and discussion}). The main conclusions of this work are summarized in section \ref{sec::conclusion}.

\section{Problem definition}
 \label{sec:: problem formulation}
\subsection{Problem definition}
We consider the sedimentation of a single spherical particle subjected to a linear cross shear flow in an EVP fluid with slight inertia. The reference frame is attached to the particle with its origin at the particle center. Thus, the particle remains stationary and the fluid moves with uniform velocity $U_{\infty}$ in the opposite direction of gravity (Figure \ref{fig::fig1}). The top and bottom walls are translating with constant and equal velocity, but in the opposite direction to create the linear shear flow in the $xz$ plane. Moreover, the walls move with the uniform velocity $U_{\infty}$ in the streamwise direction $y$. Consequently, the upper and lower plates are moving obliquely. While remaining fixed, the particle rotates freely due to the applied cross shear flow.
\subsection{Mechanical model of EVP material}
\begin{figure}
	\centering
	\includegraphics[trim={0cm 0cm 0cm 0cm},width=0.6\linewidth]{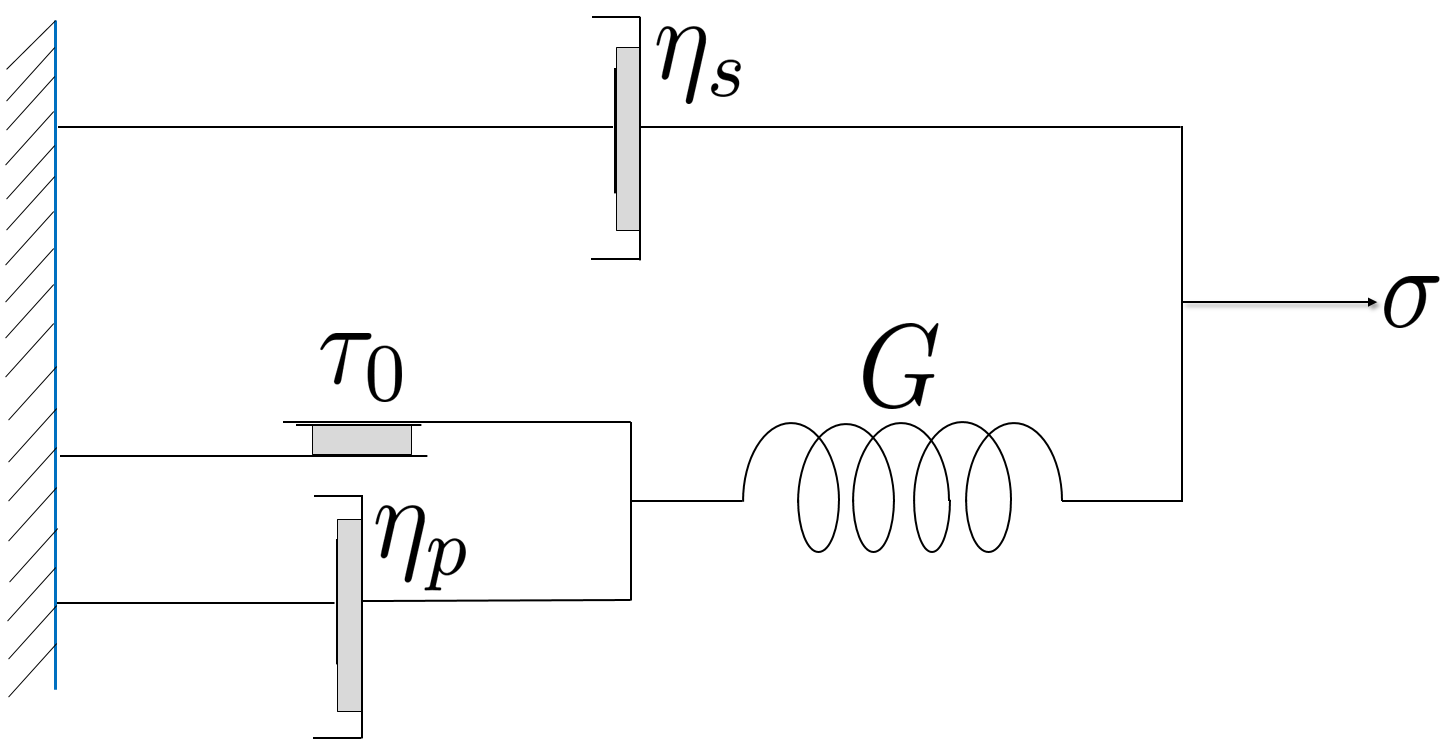} 	
	\caption{Mechanical model schematic of an EVP material proposed by \cite{saramito2009new}.}
	\label{fig::fig2}
\end{figure}

In this paper, we have adopted the model proposed by \cite{saramito2009new} to simulate the EVP material. Figure \ref{fig::fig2} represents a mechanical description of this model, based on the friction $\tau_{0}$, spring stiffness \rev{$G$} and solvent  \rev{$\eta_{s}$ and polymer $\eta_{p}$} viscosities. When the elastic strain energy of the system exceeds the threshold value (yield value), the friction element breaks allowing viscous deformation. After yielding, the deformation is unbounded in time, thus the system behaves as a non-linear viscoelastic fluid. Before yielding, however, the material behaves as a non-linear viscoelastic solid.

\subsection{Mathematical model}
This section presents the governing equations in dimensionless form. Here, the characteristic velocity scale is the uniform inflow velocity $U_{\infty}$, the characteristic length scale is the particle diameter $D$, the characteristic time scale is $D/U_{\infty}$ (since the uniform flow is the main flow), and $\eta_{0}U_{\infty}/D$ is the characteristic stress. $\eta_{0}$ is the total viscosity at zero shear-rate which is sum of the solvent $\eta_{s}$ and polymer $\eta_{p}$ viscosities $\eta_{0}=\eta_{s}+\eta_{p}$. The continuity and momentum equations are as follows:
\begin{equation}\label{mass}
\nabla. \boldsymbol{u}=0,
\end{equation}

\begin{equation}\label{mom}
Re_{p}(\frac{\partial \boldsymbol{u}}{\partial t}+\boldsymbol{u}\nabla \boldsymbol{u})=\nabla.(-pI+2(1-\beta)D(\boldsymbol{u})+\boldsymbol{\tau})+\boldsymbol{f},
\end{equation}
where $\boldsymbol{u}$ is the fluid velocity vector and $D(\boldsymbol{u})$ is the rate of deformation tensor defined as $D(\boldsymbol{u})=\frac{1}{2}(\nabla \boldsymbol{u}+ \nabla \boldsymbol{u}^{T})$. It is noteworthy to mention that, we subtract the static pressure of the fluid and scale the remainder with the viscous stress scale. Therefore, the dimensionless pressure $p$ is the dimensionless dynamic pressure. Here, $Re_{p}$ is the particle Reynolds number that is defined as the ratio of the inertial forces to viscous forces; $Re_{p}=\frac{\rho_{f}U_{\infty}D}{\eta_{0}}$, where $\rho_{f}$ is the fluid density. The retardation parameter, $\beta$, is the ratio between the polymer and the total viscosities $\eta_{p}/\eta_{0}$, $\boldsymbol{f}$ is an external body force (IB force) used to model the presence of the particle. $\boldsymbol{\tau}$ is the extra stress tensor due to the plasticity and elasticity of the EVP fluid.
\par 
The following constitutive equation is proposed to model the stress-deformation of EVP materials \cite[][]{saramito2009new}:
\begin{equation}\label{evp}
Wi \overset{\nabla}{\boldsymbol{\tau}}+\kappa_{n}(|\tau_{d}|)\alpha\boldsymbol{\tau}-2\beta D(\boldsymbol{u})=0,
\end{equation}
where $Wi$ is the shear Weissenberg number, the ratio of the material relaxation time scale $\lambda$ to the cross shear flow time scale $1/\dot{\gamma_{0}}$. Therefore, $Wi=\lambda \dot{\gamma_{0}}$. Note that the material relaxation time $\lambda$ is the ratio of polymer viscosity $\eta_{p}$ to the elasticity parameter \rev{$G$}; $\lambda=\eta_{p}/\rev{G}$.  $\alpha$ is another dimensionless parameter whose magnitude determines the primary and secondary flow; in particular, the ratio of the applied shear rate $\dot{\gamma_{0}}$ to the shear rate induced by particle settling in the fluid $\dot{\gamma}_{sett}$ is denoted by $\alpha=\dot{\gamma_{0}}/\dot{\gamma}_{sett}$. The settling shear rate is the ratio of the uniform inflow velocity to the particle diameter, i.e, $\dot{\gamma}_{sett}=U_{\infty}/D$ and thus, $\alpha=\dot{\gamma_{0}}D/U_{\infty}$. In the context of EVP fluids, $\alpha$ can be viewed as the ratio between the shear Weissenberg number $Wi$ and the settling Weissenberg number, which is called \rev{``$Wi_{\infty}$"} in the present work and is equal to $\frac{\lambda U_{\infty}}{D}$.
\par 
$\overset{\nabla}{\boldsymbol{\tau}}$ is the upper convected derivative of the stress field; it is computed based on the following relation:
\begin{equation} \label{upper convected derivative}
\overset{\nabla}{\boldsymbol{\tau}}=\frac{\partial \boldsymbol{\tau}}{\partial t}+\textbf{u}. \nabla \boldsymbol{\tau}- (\nabla \textbf{u})^{T}. \boldsymbol{\tau}-\boldsymbol{\tau}.\nabla \textbf{u}.
\end{equation}
In equation (\ref{evp}) $\kappa_{n}(|\tau_{d}|)$ is the plasticity criterion, defined by the following relation:
\begin{equation}\label{plasticity_function}
\kappa_{n}(|\tau_{d}|)=max(0,\frac{|\tau_{d}|-Bi}{(2\beta)^{1-n}|\tau_{d}|^{n}})^{\frac{1}{n}},
\end{equation}
where $|\tau_{d}|$ is the second invariant of the deviatoric part of the extra stress tensor:
\begin{equation}\label{eqn::deviatoric_tensor}
\boldsymbol{\tau_{d}}=\boldsymbol{\tau}-\frac{1}{3}tr(\boldsymbol{\tau})\boldsymbol{I},
\end{equation}
with $\boldsymbol{I}$ the unit tensor and $n$ the power-law index. $Bi$ is the Bingham number, which is the ratio of yield stress to viscous stress, $Bi=\frac{\tau_{0}D}{\eta_{0}U_{\infty}}$. $\tau_{0}$ is the material yield stress. Note that throughout the paper $Bi$ is the Bingham number defined based on the particle settling shear rate. Polymer viscosity is computed via: $\eta_{p}=\kappa (\frac{U_{\infty}}{D})^{n-1}$, where $\kappa$ is the consistency parameter.
\par 
The rotational particle velocity is computed by solving the Euler equation in the body-fixed reference frame:
\begin{equation}\label{eqn:: Euler}
I_{s}\frac{d \omega_{c}}{dt}=\oint_{\partial \Omega} \boldsymbol{r} \times (\tau. \boldsymbol{n})dA.
\end{equation}
In equation \ref{eqn:: Euler}, $I_{s}$ is the moment of inertia of the particle and is equal to $\frac{2}{5}\rho_{s}V_{s}R^{2}$, where $\rho_{s}$ is the particle density, $V_{s}$ is the particle volume and $R$ is the particle radius. Here, $\boldsymbol{n}$ is the unit normal vector at the particle surface $\partial \Omega$ and $\omega_{c}$ is the particle rotational velocity. Since our simulations are performed in a body-fixed reference frame, the particle angular velocity is tracked in an inertial reference frame by adopting the rotation matrix from \cite{chen2006flow}. For more details on the transformation the reader is referred to \cite{vazquez2016three}.

\subsection{Boundary conditions}\label{sec::boundary conditions}
The computational domain is a rectangular box with length of $16D$ in the streamwise $y$ direction, $6D$ in the spanwise $x$ direction and $12D$ in the wall-normal $z$ direction. The particle is located in the middle of the box with its center as the origin of the coordinate system. The inlet velocity boundary condition is a combination of uniform velocity and the simple shear flow. Hence, at the inlet, the fluid velocity has a component in both the streamwise and spanwise directions. The convective outflow boundary condition is applied at the outlet where the fluid is leaving the domain in the $y$ direction. The velocity is extrapolated by solving the convective equation at the location of the outlet \cite[][]{uhlmann2003first}:  
\begin{equation} \label{eqn:: outlet BC}
\frac{\partial \boldsymbol{u}}{\partial t}+U_{\infty}\frac{\partial \boldsymbol{u}}{\partial \boldsymbol{n}}=0.
\end{equation}
In order to maintain the overall mass balance and satisfy the compatibility condition, the convective velocity should be equal to the uniform inflow velocity $U_{\infty}$. A no slip boundary condition is applied at the top and bottom plates with velocities of the plates equal to $\boldsymbol{u}=-\dot{\gamma}_{0}L_{z}\hat{x}+U_{\infty}\hat{y}$ and  $\boldsymbol{u}=\dot{\gamma}_{0}L_{z}\hat{x}+U_{\infty}\hat{y}$. A homogeneous Neumann condition is used for pressure at the two walls as well as the inlet and outlet ($\partial p/ \partial \boldsymbol{n}=0$). A no flux condition is specified normal to the confinement walls for the extra stress tensor. A periodic boundary condition is applied for the velocity, pressure and extra stress tensor in the spanwise $x$ direction. The no-slip/no-penetration boundary condition is satisfied at the sphere surface implicitly by using the multidirect forcing Immersed-Boundary scheme \cite[][]{breugem2012second}. 
\par 
We analytically solve the EVP constitutive equations of \cite{saramito2009new} for the combination of Couette and uniform flow at steady state in the absence of the spherical particle and then apply this solution for the EVP stress tensor at the inlet. At steady-state the flow of EVP fluid is entirely yielded. Thus, the plasticity criteria function is positive and non-zero ($\kappa_{n}(|\tau_{d}|)>0$). In this flow configuration, all the components of the extra stress tensor are zero except the normal and shear stress in the spanwise direction of the shear plane $xz$. The non-vanishing value of the normal stress difference, which is observed in practical yield stress fluids \cite[e.g.][]{janiaud2005foam}, is the result of incorporating the elastic stresses in the EVP constitutive equation as the classical visco-plastic models \cite[such as][]{bingham1922fluidity,herschel1926measurement} do not predict the first normal stress difference in shear flows. The analytical solution is obtained by solving the following coupled non-linear equations for the normal and shear EVP stresses:
\begin{equation} \label{eqn:EVP_analytical1}
(\frac{|\tau_{d}|-Bi}{|\tau_{d}|}) \tau_{xz}=2 \alpha \beta,
\end{equation}
\begin{equation}\label{eqn:EVP_analytical2}
\tau_{xz}=\sqrt{\frac{\alpha \beta}{2Wi} \tau_{xx}},
\end{equation}
where $|\tau_{d}|=\sqrt{\frac{1}{3} \tau_{xx}^2 + \tau_{xz}^2}$ as $\tau_{xx}$ and $\tau_{xz}$ are the only non-zero elements of the stress tensor. The non-linear system of equations (\ref{eqn:EVP_analytical1}) and (\ref{eqn:EVP_analytical2}) is easily solved using the \textit{fsolve} routine of MATLAB. For the pure sedimentation simulations, the stress tensor at the inlet and outlet is set equal to zero.
\subsection{Computational matrix}
In this study, all the simulations are conducted at $\alpha=0.1$, i.e, the particle settling shear rate is ten times larger than the externally imposed shear rate. We will present a series of simulations for the problem of single sphere settling through EVP fluid in the absence and presence of simple cross shear flow. The dimensionless parameters used for the simulations presented here are reported in table \ref{table::parameters}.

\begin{table}
	\begin{center}
		\def~{\hphantom{0}}
		\begin{tabular}{lcccccc}
		Shear-induced sedimentation & \rev{$Wi_{\infty}$} & $Wi$ & $\beta$ & n & $Re_{p}$  & $Bi$ \\[3pt]
			
			& 0.1,1 & 0.01,0.1 & 0.8 & 1 & 1 & 0.05\\
			&  &  &  &  &  & 0.1\\
	     	&  &  &  &  &  & 0.13\\
	    	&  &  &  &  &  & 0.5\\
	    	&  &  &  &  &  & 1\\
	    Pure sedimentation & 0.1 & 0 & 0.8 & 1 & 1 & 0.05\\
	    	&  &  &  &  &  & 0.1\\
	    	&  &  &  &  &  & 0.13\\
	    	&  &  &  &  &  & 0.5\\
	    	&  &  &  &  &  & 1\\
		\end{tabular}
		\caption{Computational matrix} \label{table::parameters}
	\end{center}
\end{table}

\subsection{Numerical Method}
The comprehensive details of the numerical algorithm are explained by \cite{izbassarov2018computational}. In brief, the equations are solved on a Cartesian, staggered, uniform grid with velocities located on the cell faces and all the other variables (pressure, stress and material component properties) at the cell centers. All the spatial derivatives are approximated with second-order centered finite differences except for the advection terms in equation (\ref{upper convected derivative}) where the fifth-order WENO (weighted essentially non-oscillatory) scheme is adopted \cite[][]{shu2009high}. The time integration is performed with a fractional-step method \cite[][]{kim1985application}, where all the terms in the evolution equations are advanced in time with a third-order explicit Runge-Kutta scheme except for the EVP stress terms which are advanced with the Crank–Nicolson method. Moreover, a Fast Poisson Solver is used to enforce the condition of zero divergence for the velocity field. The coupling of the fluid and particles is performed with the immersed boundary method proposed by \cite{breugem2012second}.
\par 
We have adopted a grid resolution of $32$ Eulerian grid points for particle diameter. Each simulation is performed on $128$ cores working in parallel and the steady-state solution is obtained after approximately $16$ weeks, corresponding to approximately $2600$ CPU hours.

\subsection{Code validation}
The present three-dimensional numerical solver has been used and extensively validated in the past for particulate flows \cite[][]{lashgari2014laminar}, non-Netwonian flows \cite[][]{rosti2017numerical,alghalibi2018interface,rosti2018turbulent,shahmardi2019turbulent}, and multiphase problems in non-Newtonian fluids \cite[][]{de2018elastoviscoplastic}. In particular, the code is recently validated for suspensions of rigid and soft particles and droplets in EVP and viscoelastic fluids \cite[][]{izbassarov2018computational}. Nonetheless, we report here a further validation case for a simple shear flow of EVP fluid at different power law indexes $n$. 
\subsubsection{EVP Couette flow}
We consider the two-dimensional shear flow (Couette flow) of an EVP fluid. Initially, the material is at rest and starts flowing due to an externally applied constant shear rate $\dot{\gamma}_0$. The objective here is to track the time evolution of the shear stress ($\tau_{12}$, with $1$ and $2$ being the streamwise and wall-normal directions respectively) for different values of the power-law index $n$ and compare them with the analytical solution provided by \cite{saramito2009new}.
\par
\begin{figure}
	\centering
	\includegraphics[trim={0cm 0cm 0cm 0cm},width=0.55\linewidth]{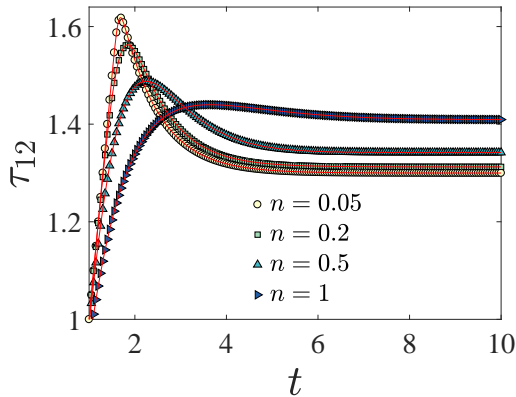} 	
	\caption{Time evolution of shear stress in a stationary simple Couette flow. The solid lines and symbols represent the analytical solution given by \cite{saramito2009new} and the present numerical simulations, respectively.}
	\label{fig::fig3}
\end{figure}

The shear Weissenberg number is fixed to $Wi=1$, the Bingham number $Bi=1$ and the retardation parameter $\beta=1$. The simulations are performed for the power-law index range $0.05\leq n \leq 1$. The time evolution of the shear stress $\tau_{12}$ and the comparison with \cite{saramito2009new} are depicted in figure \ref{fig::fig3}. We observe that, initially, the shear stress grows linearly, but as soon as the stress level achieves a threshold value, which is called ``yield stress", the growth stops and a plateau is reached; this is expected in the yielded state. Indeed as soon as the second invariant of the deviatoric stress tensor exceeds the material yield stress (or $Bi$ number in dimensionless context), the plasticity criteria function $\kappa_{n}(|\tau_{d}|)$ becomes positive and non-zero in the constitutive equation precluding the stress growth. Consequently, the solution remains bounded for any positive value of the power-law index. As shown in the figure, we find a very good agreement with the results by \cite{saramito2009new}.
\section{Results and discussion} \label{sec:: results and discussion}
In this section, we first present the results of the pure sedimentation simulations, i.e, in the absence of any external shear flow, and then those of the shear-induced sedimentation, i.e, from the simulations involving cross shear flow. 
\subsection{Pure sedimentation in EVP fluids} \label{sec:Tsam_matrix}
For the case of a sphere settling in an EVP fluid in the absence of cross shear flow, the dimensionless numbers have been chosen in such a way that the gravity number (related to Bingham number; $Y_{g}$, see equation \ref{eqn::gravity_number}) remains smaller than its critical value $Y_{g}^{c}$ reported by \cite{fraggedakis2016yielding}. Above the critical condition the yield stress resistance overcomes the buoyancy force and, consequently, the particle is trapped inside the yield-sress fluid. The critical gravity number is found to be $Y_{g}^{c}=0.143$ for Bingham fluids (i.e.\ in the absence of elasticity) by \cite{beris1985creeping}. However, numerical simulations have demonstrated that this value increases with the material elasticity.
\par
Here, we show that  the gravity numbers for the current simulations are smaller than the critical gravity number  presented as  the stoppage criteria by \cite{fraggedakis2016yielding}. To this end, we must first rescale our parameters following their definition. Thus, for each simulation we must find the corresponding Deborah number, $De$, and gravity number, $Y_{g}$, defined as:
\begin{equation}\label{eqn::gravity_number}
De=\frac{\lambda \Delta \rho g R} {\eta_{p}}; \quad Y_{g}=1.5 \frac{\tau_{0}}{\Delta \rho g R},
\end{equation}
where $\Delta \rho$ is the density difference between the bead and the fluid. The critical gravity number is obtained using non-linear regression on the simulation data in \cite{fraggedakis2016yielding}:
\begin{equation}\label{critical_gravity_Tsam}
\frac{1}{Y_{g}^{c}}=1.2+\frac{1}{0.176+0.135De}.
\end{equation}
The parameters of the simulations, including the Deborah and gravity numbers, and the comparison with the critical gravity number predicted by equation (\ref{critical_gravity_Tsam}) are shown in table \ref{table::1}. According to this table, the gravity number in the present work is always smaller than its critical value.
\begin{table}
	\begin{center}
		\def~{\hphantom{0}}
		\begin{tabular}{lcccc}
			\rev{$Wi_{\infty}$}  & $Bi$ & $De$ & $Y_{g}$ & $Y_{g}^{c}$ (eqn: \ref{critical_gravity_Tsam}) \\[3pt]
			0.1 & 0.05 & 1.31 & 0.0071 & 0.1426\\
			& 0.1 & 1.46 & 0.0128 &  0.1554 \\ 
			& 0.13 & 1.53 & 0.0159 & 0.1614 \\ 
			& 0.5 & 2.07 & 0.0453 & 0.2017 \\ 
			& 1 & 2.47 & 0.0758 & 0.2289 \\
		\end{tabular}
		\caption{Dimensionless numbers of the pure sedimentation simulations with associated gravity number $Y_{g}$, smaller than its critical value $Y_{g}^{c}$.} \label{table::1}
	\end{center}
\end{table}
\par 
\subsubsection{Velocity and stress fields}
In this section the velocity and stress fields around a sphere are demonstrated to show the effects of elasticity and plasticity on the flow dynamics. To better highlight the elasticity effects on the flow features and the sphere drag, we have performed the additional simulations of particle settling in a Bingham fluid, whose details are presented in Appendix \ref{App:Bingham}.

 Figure \ref{fig::vel_vp_EVP} shows the contour plot of the velocity magnitude in the mid $yz$ plane around the surface of the particle settling in an otherwise quiescent EVP fluid. An analogous map for the Bingham fluid is included for the sake of comparison (see methods and details in Appendix \ref{App:Bingham}). The $Bi$ number is held fixed and equal to $0.5$ for both cases and the settling Weissenberg number \rev{$Wi_{\infty}=0.1$} for the EVP material and $0$ for the Bingham fluid. The magnitude of the velocity is normalized by the particle settling velocity, $u_{sett}$. Note that we depict the flow only in a small window around the particle to highlight the shape of the velocity contours.

\begin{figure}
	\centering
	\includegraphics[trim={0cm 0cm 0cm 0cm},width=0.85\linewidth]{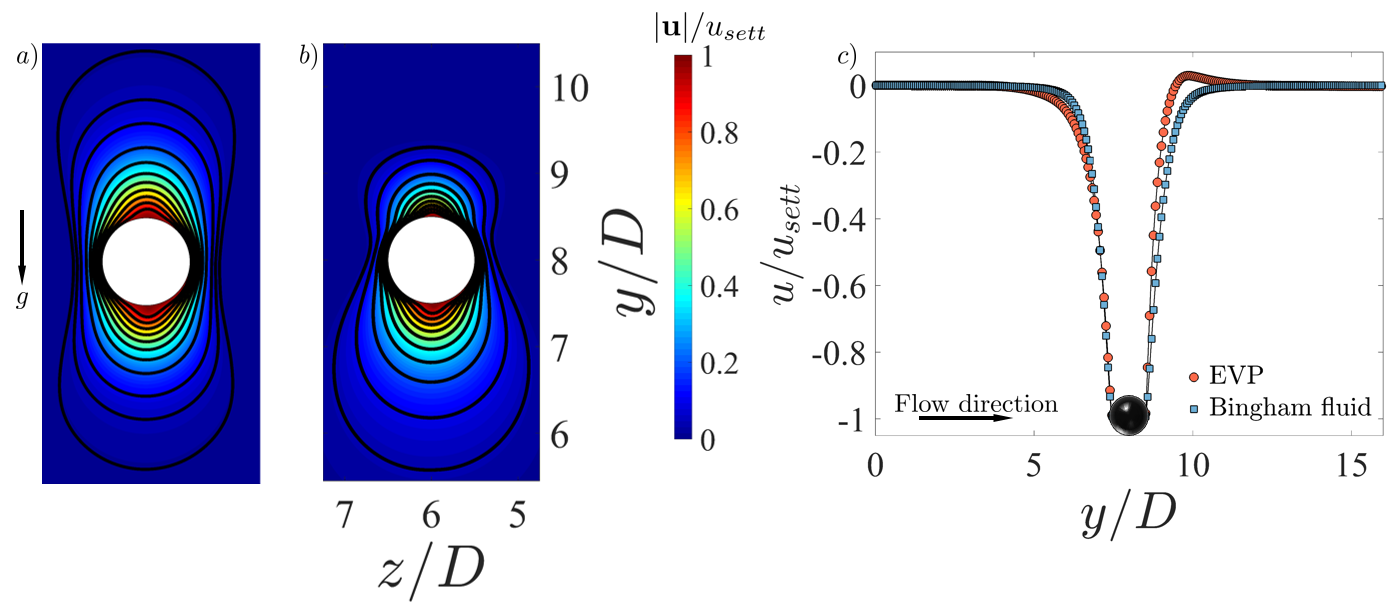} 	
	\caption{$a)$ Normalized velocity magnitude around the particle settling in the $yz$ centerplane ($x=3D$) through Bingham fluid and $b)$ EVP material at \rev{$Wi_{\infty}=0.1$}. $c)$ Streamwise velocity around a particle settling in a Bingham fluid and an EVP material  at the channel center line ($x=3D$, $z=6D$). In all figures, the Bingham number $Bi=0.5$.}
	\label{fig::vel_vp_EVP}
\end{figure}

We observe \rev{that,} the symmetry of the velocity field around the particle stagnation points (north and south poles of the sphere) is broken and an overshoot in the downstream velocity appears as soon as a small elasticity (\rev{$Wi_{\infty}=0.1$}) is added to the carrying fluid. This behavior is quantitatively depicted in figure \ref{fig::vel_vp_EVP}$c)$, where the normalized streamwise velocity is plotted in the flow direction at the center line of the channel, i.e, as function of $y$ for $x=3D$, $z=6D$.

\par 
The loss of the fore-aft symmetry in a yield stress fluid and the negative wake formation have been previously observed in experiments  
\cite[][]{gueslin2006flow,putz2008settling,holenberg2012particle,ahonguio2014influence}. 
 \cite{fraggedakis2016yielding} revealed that the elasticity of realistic yield stress fluids is the primary cause of such phenomena. Indeed, these authors  demonstrated that the thixotropy (aging of yield stress materials) is not responsible for such behavior as conjectured before. The absence of this symmetry was also observed in the case of a neutrally-buoyant sphere in Carbopol gels subjected to simple shear flow \cite[][]{firouznia2018interaction}. Therefore, to correctly predict the behavior of practical yield stress fluids, the effect of elasticity should be taken into account.
It is also worthwhile to mention that the loss of the fore-aft symmetry is not due to the weak inertia of our simulations ($Re_{p}=1$), see section \ref{sec:drag_pure_sedimentation} for an explanation.
\par
\begin{figure}
	\centering
	\includegraphics[trim={0cm 0cm 0cm 0cm},width=0.7\linewidth]{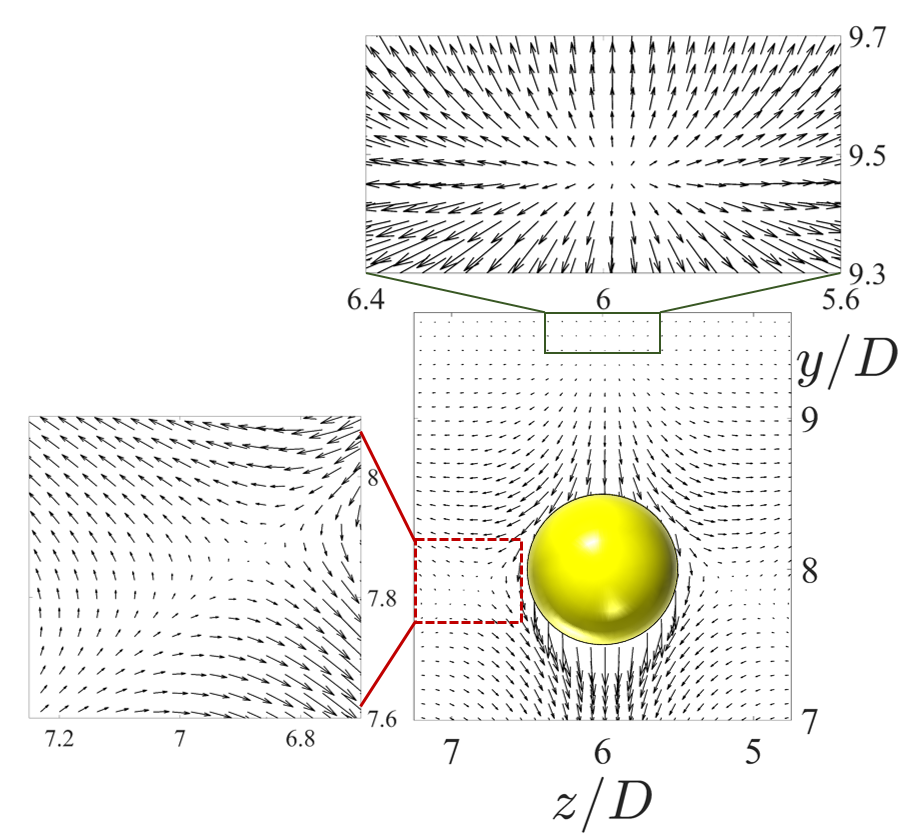} 	
	\caption{Velocity vectors around the sphere settling in an EVP material in the central $yz$ plane. The dashed red and solid green boxes magnify the recirculation zones and the flow stagnation points. The dimensionless numbers are the same as figure \ref{fig::vel_vp_EVP}.}
	\label{fig::vel_vector}
\end{figure}

 
The velocity vector field around the particle settling in an EVP fluid is illustrated in figure \ref{fig::vel_vector} for the same case as in figure \ref{fig::vel_vp_EVP}.
For clarity, the recirculation zones (inside of the dashed red box) and flow stagnation points (from the solid green box) are magnified. 
The negative wake is observed where the fluid velocity is opposite the particle velocity, upstream of the rear stagnation point. The overshoot in the stream velocity downstream of the particle shown in figure \ref{fig::vel_vp_EVP}$c)$ occurs in the same area.
For a particle with no-slip boundary condition, the recirculation zones arise close to the particle and in its equatorial plane as previously observed experimentally \cite[][]{holenberg2012particle}, which could also be captured computationally by adopting an extensive mesh refinement near the particle surface \citep{fraggedakis2016yielding}. 
\par
\begin{figure}
	\centering
	\includegraphics[trim={0cm 0cm 0cm 0cm},width=0.95\linewidth]{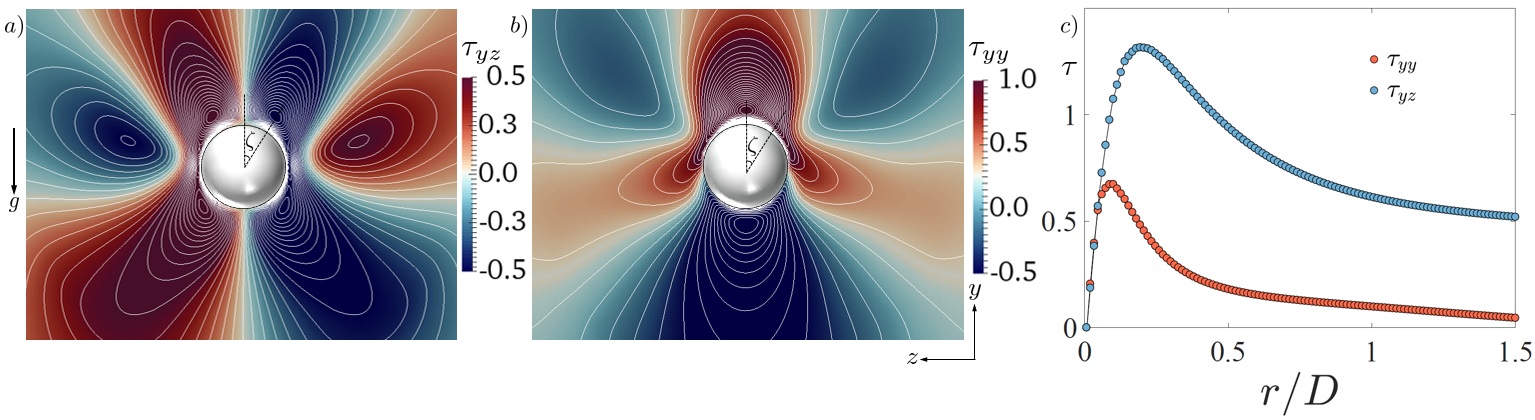} 	
	\caption{Stress fields around the particle settling in an EVP fluid in the $yz$ plane at $x=3D$ $a)$ the $\tau_{yz}$ shear stress  and $b)$ the $\tau_{yy}$  normal stress. The angle $\zeta$ is measured from the north pole of the sphere. $c)$ Stress relaxation versus the distance from the particle, $r$, at an angle $\zeta=30^{\circ}$ measured from the north pole of the sphere. $r=0$ corresponds to the particle surface. The absolute value of the stresses are shown for the same case as in figure \ref{fig::vel_vp_EVP}.}
	\label{fig::stress_sed}
\end{figure}

 
The interaction between shear and normal stresses upstream of the particle is shown to be the main cause for the negative wake formation \citep{fraggedakis2016yielding}. Infact, the negative wake is due to the normal stress relaxing faster than the shear stress away from the particle. This is shown in the colormap of normal and shear stresses around the sphere in figures \ref{fig::stress_sed}$a)$ and $b)$. The stress relaxation away from the sphere, in particular the shear and normal stresses, are also depicted along the ray at an angle $\zeta=30^{\circ}$, measured from the rear stagnation point (sphere north pole) in figure \ref{fig::stress_sed}$c)$ (we define the north pole in the positive $y$ direction).

Clearly, at $r\approx 0.2D$, the absolute value of the shear stress is twice as large as the normal stress. Thus, the elastic shear stress, which is responsible for the creation of a rotational force on the fluid element, contributes to the negative wake and to the secondary flow upstream of the particle. For more details the reader is referred to \cite{fraggedakis2016yielding}.
\par

\subsubsection{Yielded/unyielded regions}
\begin{figure}
	\centering
	\includegraphics[trim={0cm 0cm 0cm 0cm},width=0.8\linewidth]{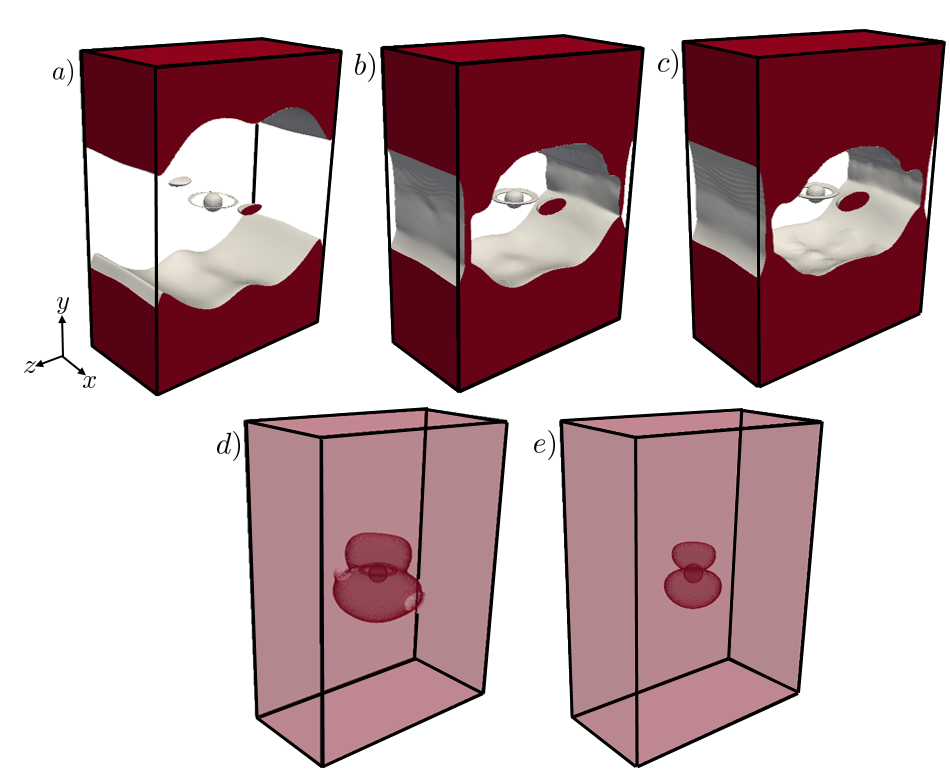} 	
	\caption{Surfaces of unyielded regions around the particle in an EVP material at \rev{$Wi_{\infty}=0.1$} in the absence of cross shear flow for various $Bi$; $a) Bi=0.05$  $b) Bi=0.1$ $c)Bi=0.13$ $d)Bi=0.5$ $e)Bi=1$. Red represents the unyielded zone, while gray shows the yield surface boundary.}
	\label{fig::fig4}
\end{figure}

\begin{figure}
	\centering
	\includegraphics[trim={0cm 0cm 0cm 0cm},width=0.8\linewidth]{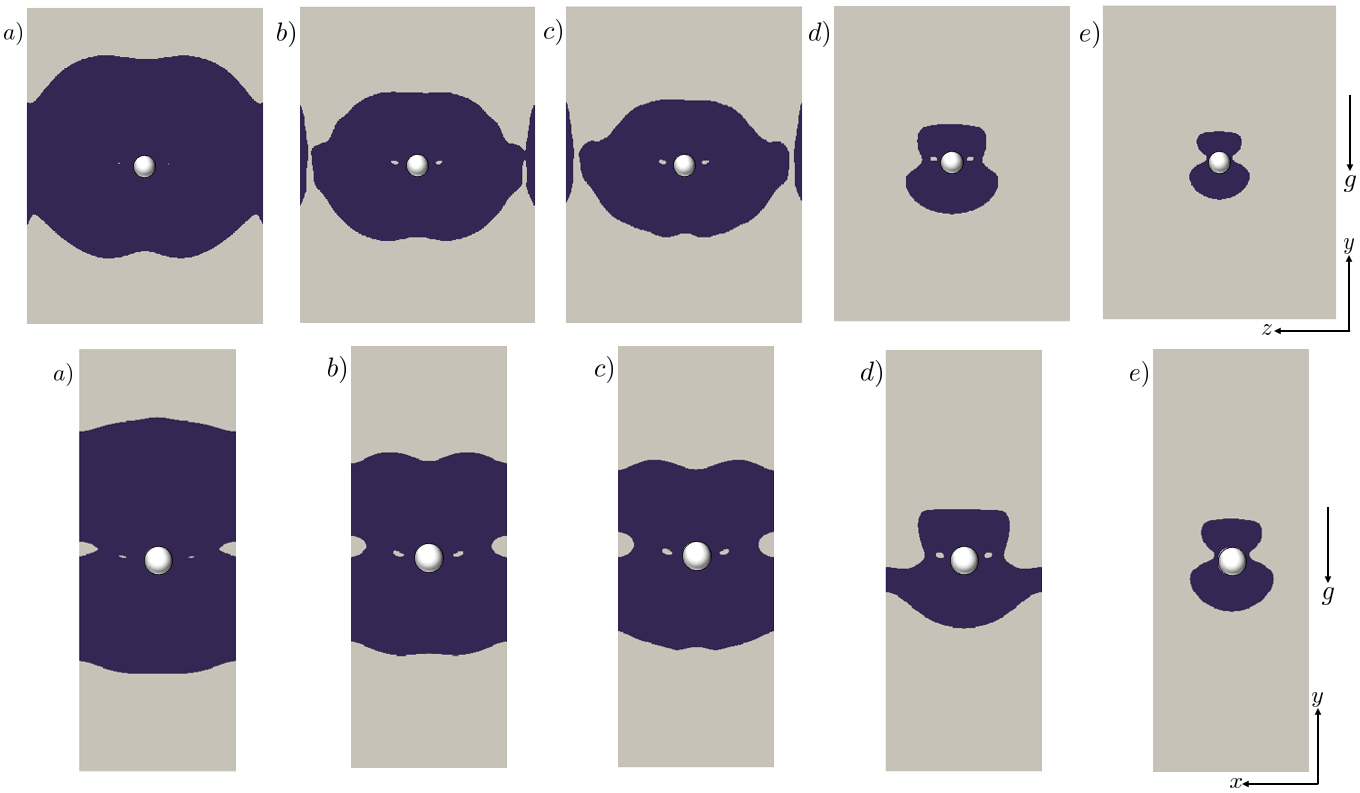} 	
	\caption{Evolution of the unyielded zones for flow of an EVP material around a particle for various $Bi$ numbers at \rev{$Wi_{\infty}=0.1$}; $a) Bi=0.05$  $b) Bi=0.1$ $c)Bi=0.13$ $d)Bi=0.5$ $e)Bi=1$. The first and second row represents the unyielded zones in the central $yz$ and $xy$ planes respectively. Blue and gray represent the yielded and unyielded regions.}
	\label{fig::fig5}
\end{figure}

In this section, we show the shape and extent of the yielded/unyielded zones around a sphere at different $Bi$ numbers in an EVP material. 
To our knowledge, the unyielded regions have not been previously reported in $3D$. 
Figure \ref{fig::fig4} displays the surfaces delimiting the yieldied regions at different $Bi$ numbers and constant elasticity (\rev{$Wi_{\infty}=0.1$}), where unyielded surfaces are plotted with lower transparency for higher $Bi$ numbers ($Bi=0.5,1$) for the sake of clarity. 
In the figure, red represents the unyielded regions, while gray depicts the yielded surface boundary and the particle. For completeness, figure \ref{fig::fig5} shows the projection of unyielded surfaces onto the mid $yz$ and $xy$ planes. In this figure, blue and gray colors denote the yielded and unyielded zones respectively.

\begin{figure}
	\centering
	\includegraphics[trim={0cm 0cm 0cm 0cm},width=1\linewidth]{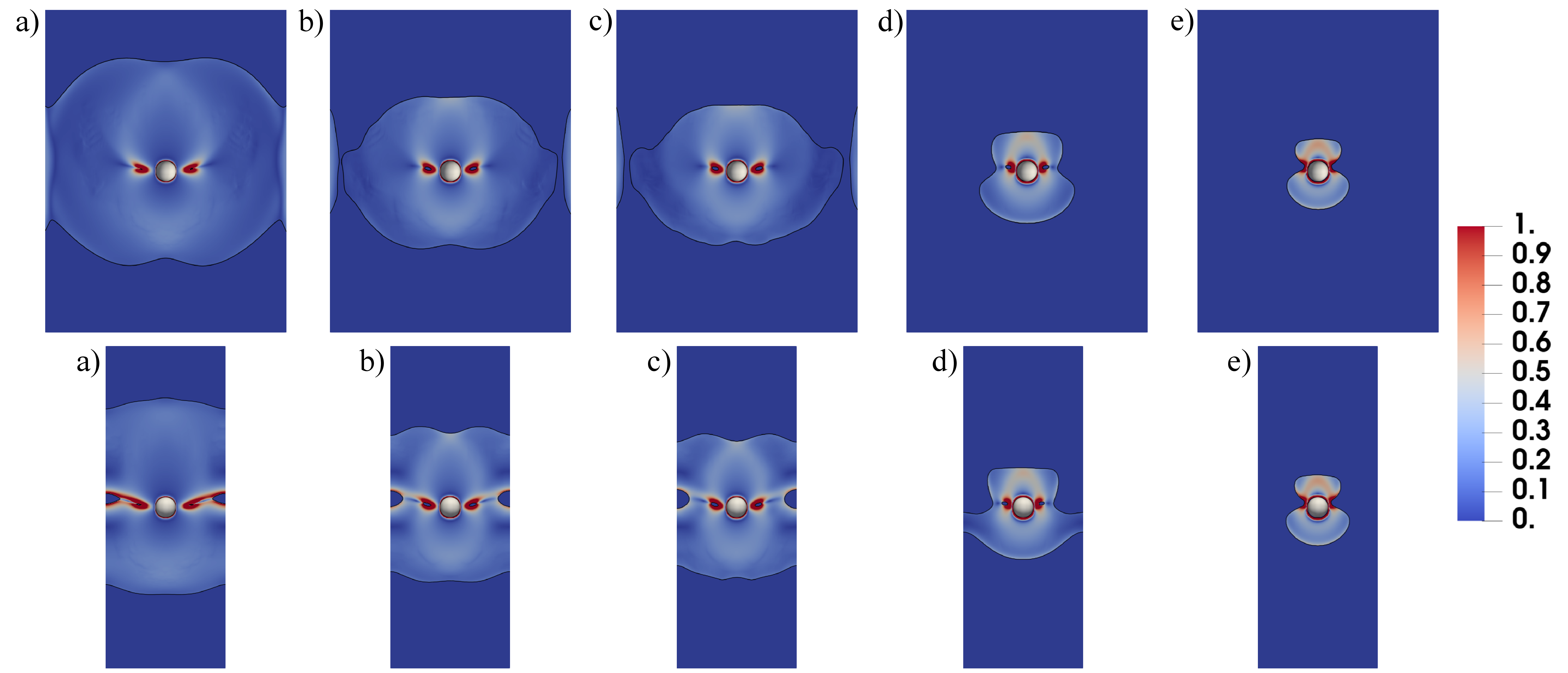} 	
	\caption{\rev{Colormaps of the second invariant of the shear rate for the flow of an EVP material around a particle for various $Bi$ numbers at $Wi_{\infty}=0.1$; a) $Bi=0.05$  b) $Bi=0.1$ c) $Bi=0.13$ d) $Bi=0.5$ e) $Bi=1$. The first and second row represents the shear rate in the central $yz$ and $xy$ planes, respectively.}
	\label{fig::App_B_Fig}}
\end{figure}

\par

\rev{The simulations reveal the existence of two unyielded zones; the first is the unyielded envelope that surrounds the fluid zone; the second is the unyielded ring located in the yielded zone surrounding the sphere. The projection of this ring in the central $yz$ and $xy$ planes is shown in figure \ref{fig::fig5}. These are the two unyielded islands in the yielded region located at the equator and on either side of the sphere. It is noteworthy to mention that these solid islands  are the regions in which the second invariant of the shear rate is almost zero as shown in figure \ref{fig::App_B_Fig}. Therefore, unlike the case of 2D cylinders, these solid rings are not rotating solid islands; note also that, at steady state, the sphere rotating velocity is zero in the pure sedimentation cases. }

\par 

\rev{We note that the outer unyielded envelope  grows progressively as the $Bi$ number increases and the yield surface boundary approaches the surface of the particle from the equator plane causing the particle to stop settling. A similar arrest  mechanism has been captured previously in axisymmetric particle-settling simulations in an EVP material \citep{fraggedakis2016yielding}.}
\rev{Moreover,  figure  \ref{fig::fig5} shows that  there exists  yielded regions in the vicinity of  the channel walls for $Bi=0.1 \& 0.13$. These  are associated with the wall effects and had also  previously been observed by \cite{blackery1997creeping} in the Bingham fluid flow past a sphere contained in a tube with a diameter that is $10$ times larger than that of the sphere.}

\subsubsection{Drag coefficients} \label{sec:drag_pure_sedimentation}
In our simulations we fix the particle and compute the drag exerted on the particle by the surrounding fluid. Since the settling rate is inversely proportional to the drag, drag enhancement is equivalent to settling rate reduction \cite[][]{padhy2013simulations}. The drag coefficient $C_{d}$ is defined as follows:
\begin{equation}\label{eqn::drag_coefficient}
C_{d}=\frac{2F_{d}}{\eta_{0}U_{\infty}D},
\end{equation} 
where $F_{d}$ is the total drag force exerted on the particle in the streamwise $(y)$ direction.
\par  
To investigate the dependency of the drag change on the flow and fluid parameters, we decompose the total drag force on the particle into its individual components, which
are associated with the different stress contributions. As we assume a finite particle Reynolds number ($Re_{p}=1$), the total drag consists of four components; namely \textit{form drag} ($F_{d}^{f}$), \textit{viscous drag} ($F_{d}^{v}$), \textit{polymer drag} ($F_{d}^{p}$) and \textit{inertia drag} ($F_{d}^{I}$). The different  components are computed from the following definitions:
\begin{equation}\label{eqn:form drag}
F_{d}^{f}=-\int \hspace{-1.5ex} \int_{\partial \Omega} pn_{y}dS,
\end{equation}
\begin{equation}\label{eqn:viscous drag}
 F_{d}^{v}=(1-\beta)\int \hspace{-1.5ex} \int_{\partial \Omega} (\frac{\partial u_{y}}{\partial x_{j}}+\frac{\partial u_{j}}{\partial y}) n_{j} dS,
\end{equation}
\begin{equation}\label{eqn:polymer drag}
F_{d}^{p}=\int \hspace{-1.5ex} \int_{\partial \Omega} \tau_{yj}n_{j} dS,
\end{equation}
\begin{equation}\label{eqn:inertia drag}
F_{d}^{I}=Re_{p} \int \hspace{-1.5ex} \int_{\partial \Omega} u_{j}\frac{\partial u_{y}}{\partial x_{j}} n_{j} dS,
\end{equation}
where $\partial \Omega$ is non-dimensionalized by $D^{2}$. The details of the numerical integration procedure are given in appendix \ref{App: A}.
\par
The form drag (eq. \ref{eqn:form drag}) is the drag component resulting from the distribution of the dynamic pressure on the sphere. \rev { It is noteworthy to mention that in the Saramito's model \citep{saramito2009new} the extra elastic stress tensor is not traceless, i.e, $\tr\left(\tau\right)$ is not zero. Thus, the pressure field $p$ obtained in the numerical solution is basically the field of $\bar{p} = p + \left(\frac{1}{3} \tr \left(\tau\right) \right)$. However, in all of our simulations, the absolute value of the $\left(\frac{1}{3} \tr \tau\right)$  at the surface of the sphere is negligible when compared to the absolute value of $\bar{p} = p + \left(\frac{1}{3} \tr \tau\right)$. 
In addition, unlike $\bar{p}$, the sign of $\left(\frac{1}{3} \tr \tau\right)$ does not change  at the stagnation points. Therefore, we conclude that the contribution of  $\left(\frac{1}{3} \tr \tau\right)$ to the drag is negligible as compared to the contribution of the dynamic pressure field $p$ for  the simulations conducted  in this work.}

\par 
To significantly reduce the time of each computations, we have assumed small but finite inertia, $Re_{p}=1$, and still expect inertial effects to be negligible. To a posteriori check this, we have compared  the sum of the form, viscous and polymer drag force to the IB force $\boldsymbol{f_{j}}$, i.e.\ the total drag, and found a difference of about $1.5\%$,  which indeed indicates negligible inertial drag, $F_{d}^{I}$.

\begin{figure}
	\centering
	\includegraphics[trim={0cm 0cm 0cm 0cm},width=0.55\linewidth]{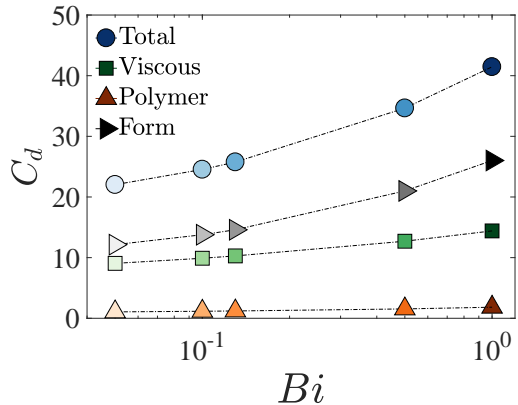} 	
	\vspace{-3ex}
	\caption{Total drag coefficient and its individual components for a  sphere settling in an EVP fluid as a function of the $Bi$ number at \rev{$Wi_{\infty}=0.1$}.}
	\label{fig::fig8}
\end{figure}
\par

Figure \ref{fig::fig8} illustrates the variation of the total drag coefficient and its individual contributions (form, viscous and polymer drag) experienced by the sphere settling in an EVP fluid as a function of the $Bi$ number. The total drag increases as the material yield stress is increased, as  
observed in the past for the particle settling in a purely visco-plastic fluid \cite[][]{beris1985creeping,atapattu1995creeping,blackery1997creeping,tabuteau2007drag,holenberg2012particle,ahonguio2014influence,wachs2016particle} as well as in an EVP fluid \cite[][]{fraggedakis2016yielding}. Here, the key finding of our 3D simulations is that the drag contribution from the dynamic pressure distribution on the particle (form drag) is dominant. Additionally, the viscous drag increases with $Bi$, whereas the drag contribution due to the polymer stresses remains almost constant with $Bi$. 

\begin{figure}
	\centering
	\includegraphics[trim={0.6cm 0cm 0cm 0cm},width=1\linewidth]{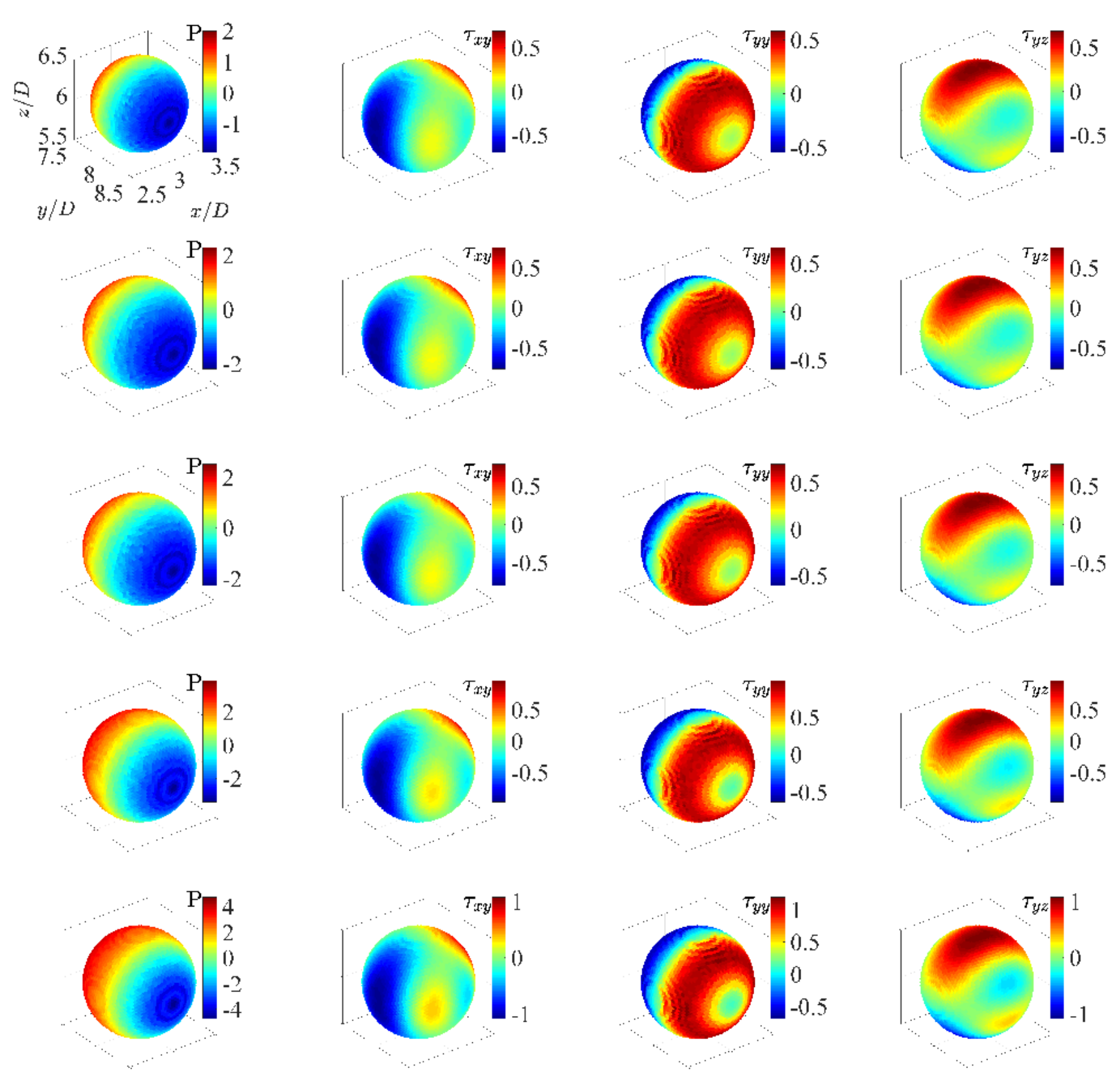} 	
	\vspace{-3ex}
	\caption{Contour plots of the dynamic pressure and the of three components of the viscous stress distribution on the surface of the particle settling in an EVP fluid at different $Bi$ numbers and \rev{$Wi_{\infty}=0.1$}. First, second, third, fourth and fifth row shows the distributions at $Bi=0.05,0.1,0.13,0.5$ and $1$ respectively.}
	\label{fig::fig9} 
\end{figure}

\par 
The dynamic pressure and of the three components of the viscous stress distribution ($\tau_{xy}, \tau_{yy}, \tau_{yz}$) are computed on the surface of the sphere and are shown in figure \ref{fig::fig9} (the computational procedure is explained in details in appendix \ref{App: A}). \rev{ Note that in this study we are dealing with a complex flow,
i.e.\
the type of flow is in general different at each point, e.g. predominantly shear, extensional, nearly rigid-body motion or a combination of these. 
However, our goal is to show the components of the stress that contribute to the drag force, i.e., $F_i=\int_{\partial \Omega} \tau_{ij}.n_j ~d A  $. For a drag force in the $y$ direction, the relevant components of the stress tensor are $\tau_{yx}, \tau_{yy}$  and $\tau_{yz}$, so we just examine these components of the viscous stress tensor,  since figure \ref{fig::fig8} shows the polymer stress contribution to the drag force is negligibile.   }

 The visualizations in figure  \ref{fig::fig9} confirm that the magnitude of the pressure on the particle surface increases with $Bi$. 
Moreover, the fore-aft symmetry in the pressure distribution around the particle stagnation points breaks leading to further drag enhancement. 
Increasing the Bingham number, on the other hand, merely magnifies the value of the shear and normal viscous stresses around the poles and on either sides of the sphere and does not break the fore-aft symmetry. Hence, in the presence of elasticity, the sphere drag increases through modification of the dynamic pressure distribution by both breaking its fore-aft symmetry with respect to its north and south poles and by magnifying its magnitude. 
\par 
Table \ref{table::VP_EVP_Sed_Drag} compares the total drag coefficient and its individual components for a sphere settling in an ideal visco-plastic (Bingham) fluid and an EVP material. Three features are evident here. First, adding a small amount of elasticity to an ideal-yield stress fluid causes a total drag reduction of approximately $20\%$, in agreement with the observations by \cite{fraggedakis2016yielding}. Second, adding elasticity to the purely visco-plastic fluid (Bingham fluid) modifies the stress fields on the sphere surface. Table \ref{table::VP_EVP_Sed_Drag} indeed shows that the dominant drag component for the Bingham fluid case is the viscous drag, while it is the form drag in an EVP fluid. Third, although the polymer drag contribution is small compared to the other components, the total drag is indirectly affected by the presence of the polymers through modification of the form and viscous stresses.

\begin{table}
	\begin{center}
		\def~{\hphantom{0}}
		\begin{tabular}{l c c}
			Drag & EVP fluid & visco-plastic (VP) fluid\\ 
			Total   & 22.03 & 27.15 \\
			Form    & 12.14 & 11.26 \\
			Viscous & 9.05 &  15.89 \\
			Polymer & 0.84 &   - \\
		\end{tabular}
		\caption{Comparison the total drag coefficient and its components on the surface of the sphere settling in a visco-plastic (VP) and in an EVP fluid at $Bi=0.05$. The settling Weissenberg number \rev{$Wi_{\infty}=0.1$} for EVP material.} \label{table::VP_EVP_Sed_Drag}
	\end{center}
\end{table}
\par 
Given the total drag, we compute the elastic contribution (elastic drag) for a sphere settling in a yield stress fluid using the following relation:
\begin{equation}\label{eqn:Elastic_drag}
(\eta_{0} U_{\infty} D) C_{d}^{EVP}=(\eta_{0} U_{\infty} D) C_{d}^{VP}- (\eta_{0} U_{\infty} D)C_{d}^{e},
\end{equation}
where $\eta_{0} U_{\infty} D$ is the viscous force scale, and $C_{d}^{EVP}$ and $C_{d}^{VP}$ the total drag coefficient on the sphere settling in an EVP material and a visco-plastic fluid. Thus, $(\eta_{0} U_{\infty} D) C_{d}^{e}$ quantifies the indirect effect of adding polymers to a yield stress fluid by modifying the dynamic pressure and viscous stresses.

\par

\rev{Note that throughout this manuscript we use the formulation ``adding polymers to a Bingham fluid" to indicate adding a finite elasticity to a Bingham fluid, although the elastic effects can be present in practical yield stress fluids  also in the absence of polymers, e.g.,  in emulsions. Moreover, in this manuscript  we use the term ``polymer drag"  to denote  the drag contribution due to the extra elastic stress tensor and the term ``elastic drag" as introduced by equation (\ref{eqn:Elastic_drag}) to indicate the total change in the drag force due to the addition of  a finite elasticity to the ideal yield stress fluids. }

\par
\begin{figure}
 	\centering
 	\includegraphics[trim={0cm 0cm 0cm 0cm},width=0.55\linewidth]{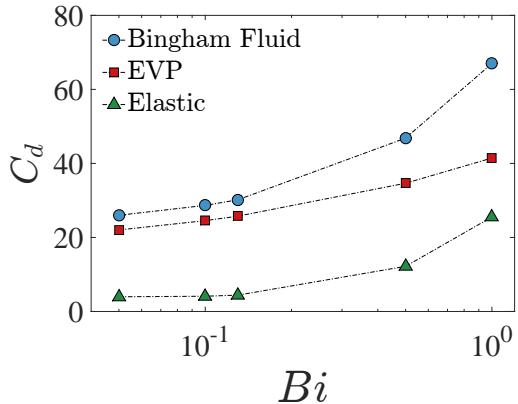} 	
 	\vspace{-2ex}
 	\caption{Total drag coefficient of a particle settling in a Bingham fluid from \cite{blackery1997creeping} and that of a particle settling in an EVP material (with elasticity \rev{$Wi_{\infty}=0.1$})  as well as the elastic drag resulting from equation \ref{eqn:Elastic_drag}.}
 	\label{fig::elastic_drag}
 \end{figure}

The variation of the total drag for a sphere settling in a Bingham fluid as a function of $Bi$ is then found using the correlation given by \cite{blackery1997creeping}, which is described in section \ref{App::Validation_Bingham}, assuming
the coefficients in equation (\ref{eqn:Stokes_drag_mitsoulis2}) $a=2.343$ and $b=0.879$ corresponding to a confinement ratio of $12$ \cite[][]{blackery1997creeping}. 
The total  drag coefficient  for settling in Bingham and EVP fluids along with the elastic drag $C_{d}^{e}$ defined above are displayed in figure  \ref{fig::elastic_drag} versus the Bingham number. Clearly, the elastic drag is an increasing function of $Bi$ for a particle settling in EVP material at constant \rev{$Wi_{\infty}$}. Moreover, the drag reduction through introduction of elasticity, which was shown in table \ref{table::VP_EVP_Sed_Drag} for a single $Bi$ number ($Bi=0.05$) applies to a range of $Bi$ numbers ($0.05 \leq Bi \leq 1$). 
\par 
We can relate the sphere drag reduction in the EVP material to the volume of the yielded region around the sphere: indeed, increasing the Deborah number at a fixed Bingham number results in an increase of the yielded region around the sphere, as previously shown by \cite{fraggedakis2016yielding}. Consequently, elasticity helps the sphere to translate faster in the EVP material by reducing the drag at high elasticity.

\par 
\begin{figure}
	\centering
	\includegraphics[trim={0.6cm 0cm 0cm 0cm},width=1\linewidth]{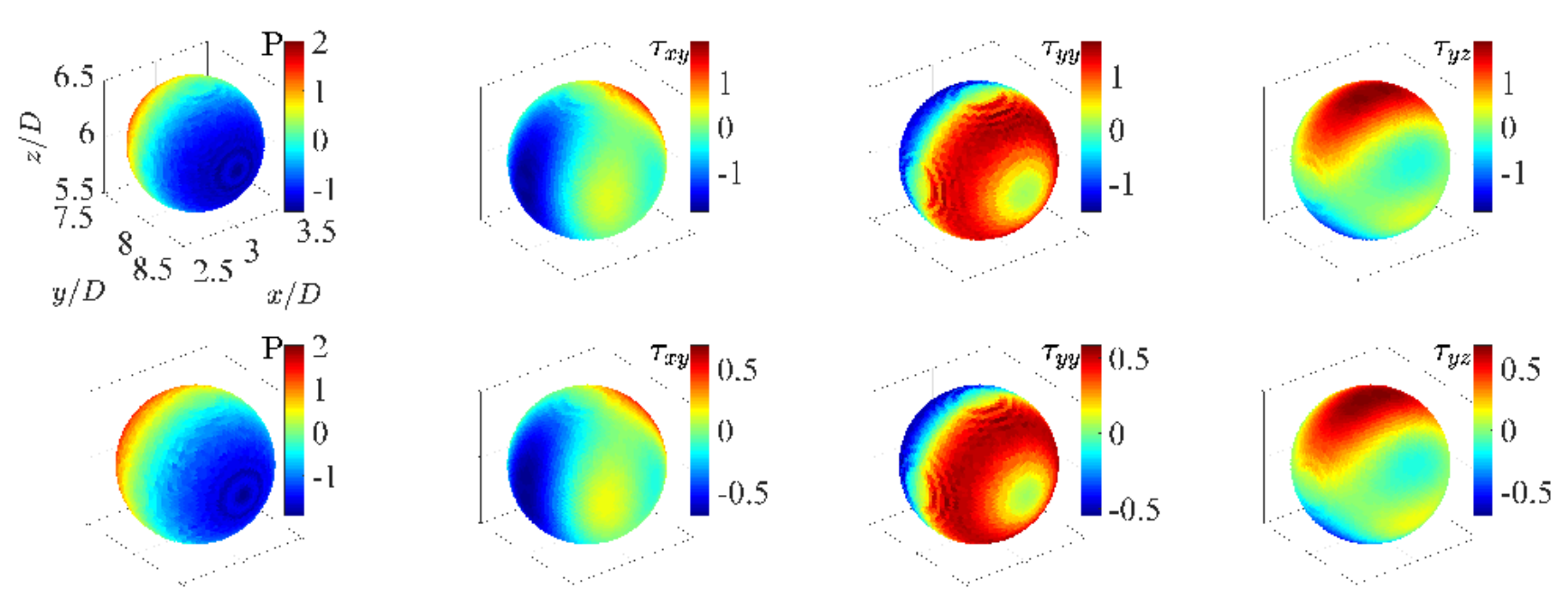} 	
	\vspace{-2ex}
	\caption{Contour plots of the dynamic pressure and of three components of viscous stress distribution on the surface of the particle settling in a Bingham fluid (first row) at $Bi=0.05$ and in the EVP fluid (second row) at $Bi=0.05$, \rev{$Wi_{\infty}=0.1$}.}
	\label{fig::fig11}
\end{figure}

%
%

To gain further insight, we display in  figure \ref{fig::fig11} the contour plots of the dynamic pressure and of the $\tau_{xy}$, $\tau_{yy}$ and $\tau_{yz}$ components of the viscous stress distributions on the sphere surface for the case of a Bingham fluid (first row) and that of an EVP fluid (second row). The pressure distribution and its magnitude are not affected by the presence of the polymers as seen in table \ref{table::VP_EVP_Sed_Drag} where the form drags are found to be close to each other. Nevertheless, the magnitudes of the viscous shear and normal stresses drop by adding polymers to the pure Bingham fluid. 
\begin{figure}
	\centering
	\includegraphics[trim={0cm 0cm 0cm 0cm},width=1\linewidth]{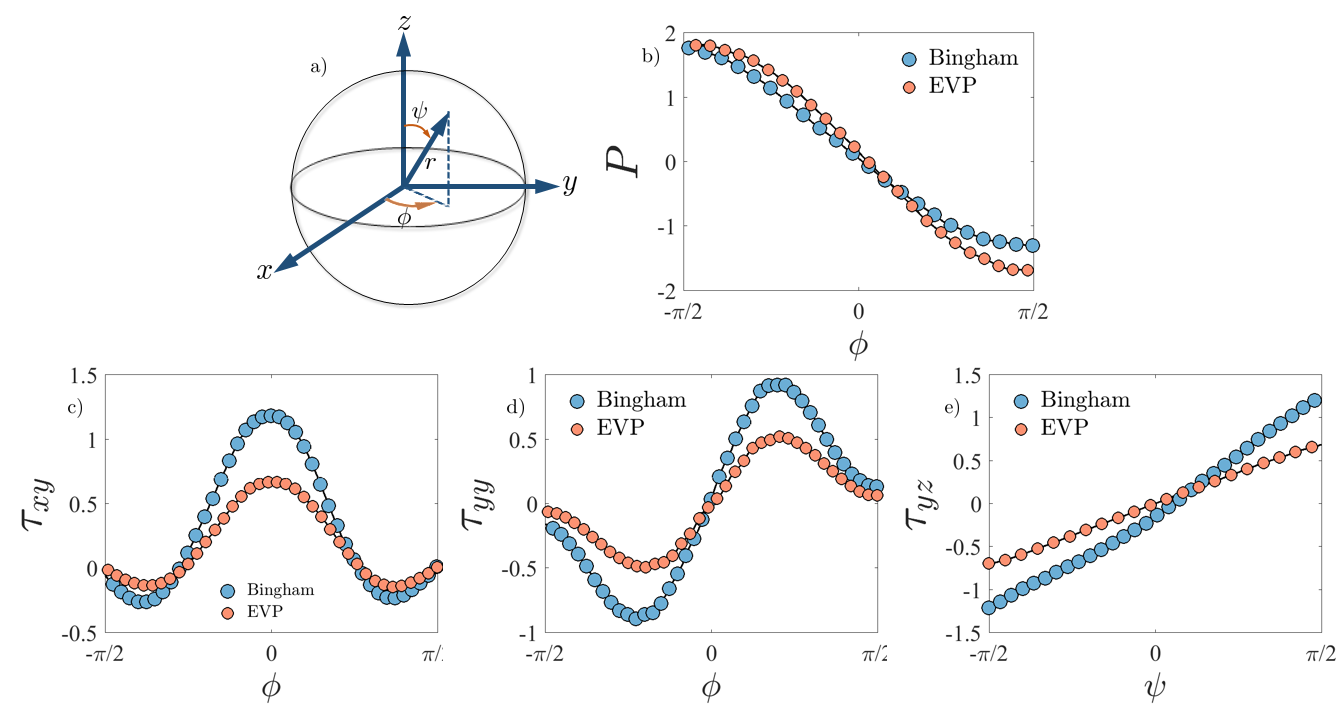} 	
	\caption{Pressure and viscous stress distributions on the surface of the sphere settling in Bingham and EVP fluids for the same cases depicted in figure \ref{fig::fig11}. $a)$ Spherical coordinate system $b)$ pressure $c)$ $\tau_{xy}$ $d)$ $\tau_{yy}$ $e)$ $\tau_{yz}$. The pressure, $\tau_{xy}$ and$\tau_{yy}$ distributions are plotted for $\psi=0$, while the $\tau_{yz}$ is plotted at $\phi=0$.}
	\label{fig::fig12}
\end{figure}
This behavior can be better understood by looking at figure \ref{fig::fig12}, where the pressure and the $\tau_{xy}$ and $\tau_{yy}$ components of the viscous stresses are plotted around the sphere stagnation points (at zero polar angle $\psi=0$ and as a function of azimuthal angle $\phi$; see coordinate system in panel a). Figure \ref{fig::fig12} also shows the $\tau_{yz}$ component of the viscous stresses on either side of the sphere at $\phi=0$ and as a function of polar angle $\psi$. Polymers affect the viscous stresses on the sphere surface, which results in total drag reduction as compared to the case of a Bingham fluid. Furthermore, the slope of the $\tau_{yz}$ contribution of the viscous stress changes in the case of an EVP fluid.

\subsection{Shear-induced sedimentation}
In this section, we present the results of the simulations of shear-induced sedimentation of a sphere in an EVP material. In our setup, the only non-zero component of the particle rotational velocity is $\omega_{y}$ owing to symmetry. The simulations are performed at a constant ratio of shear to sedimentation Weissenberg number $\alpha=\frac{Wi}{\rev{Wi_{\infty}}}=0.1$. Equivalently, the ratio between the externally imposed cross shear rate to the shear rate induced by settling is kept fixed and equal to $0.1$ ($\dot{\gamma}_{0}/\dot{\gamma}_{sett}=0.1$). This means that the cross shear flow is always a secondary flow and the uniform flow is the primary flow. 
\subsubsection{Velocity field}
\begin{figure}
	\centering
	\includegraphics[trim={0cm 0cm 0cm 0cm},width=0.9\linewidth]{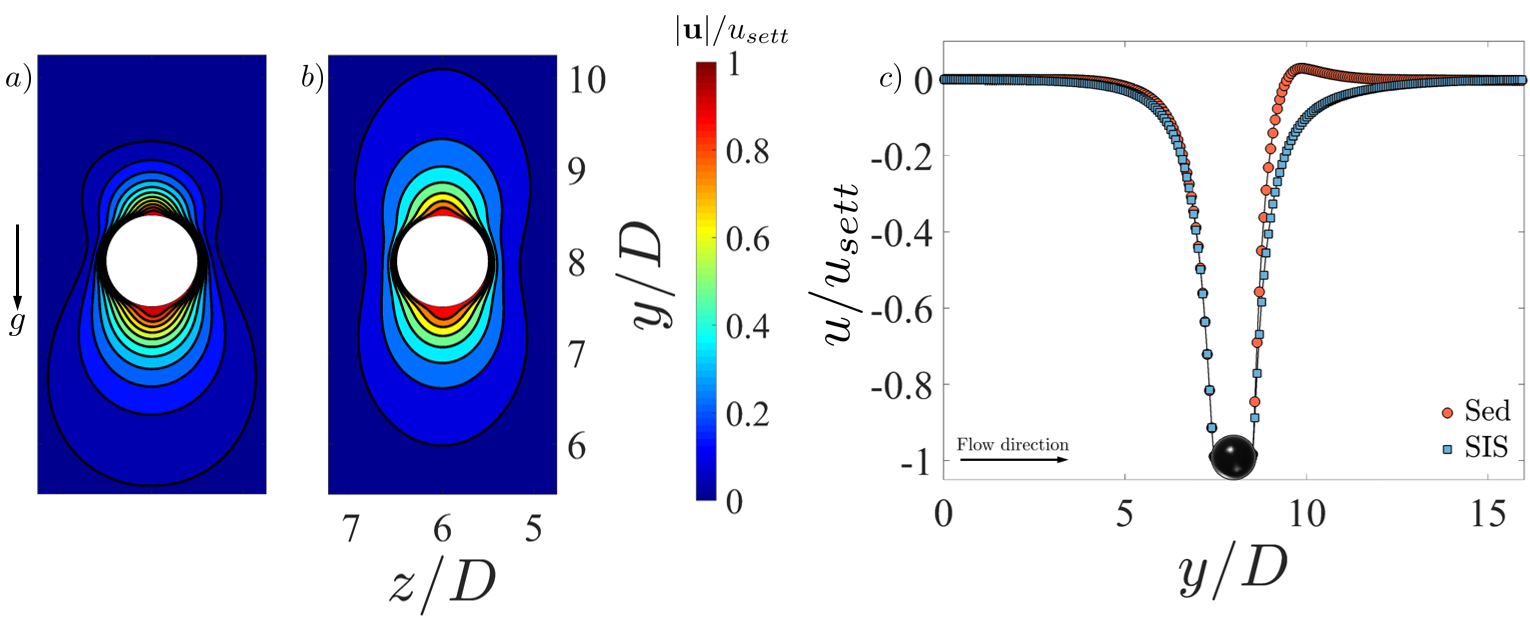} 	
	\caption{$a)$ Normalized velocity colormaps in the $yz$ centerplane ($x=3D$) for the case of pure sedimentation of a single sphere in an EVP material. $b)$ The same as $a)$, but for the case of shear-induced sedimentation. $c)$ Streamwise fluid velocity in the flow direction at the channel centerline ($x=3D$, $z=6D$) for the cases of pure sedimentation (orange markers) and shear-induced sedimentation (blue markers) in an EVP material. In all figures the dimensionless numbers are: $Bi=0.5$, \rev{$Wi_{\infty}=0.1$}.}
	\label{fig::vel_sed_SIS}
\end{figure}

The velocity field around a sphere settling in a sheared and quiescent EVP fluids is presented in this section. Figure \ref{fig::vel_sed_SIS} illustrates the velocity magnitude normalized by the particle settling velocity in the mid plane between the two walls; results are presented for settling both in the presence and absence of an externally imposed cross shear flow.
We observe that the fore-aft asymmetry of the velocity field around the sphere is less pronounced when the shear flow is superimposed on the uniform flow. This feature is further clarified in figure \ref{fig::vel_sed_SIS}$c)$, which illustrates the differences due to the presence of the cross shear flow in terms of streamwise velocity. Clearly, the velocity overshoot downstream of the sphere disappears once the secondary orthogonal cross shear flow is superimposed to the primary uniform flow.

\begin{figure}
	\centering
	\includegraphics[trim={0cm 0cm 0cm 0cm},width=0.7\linewidth]{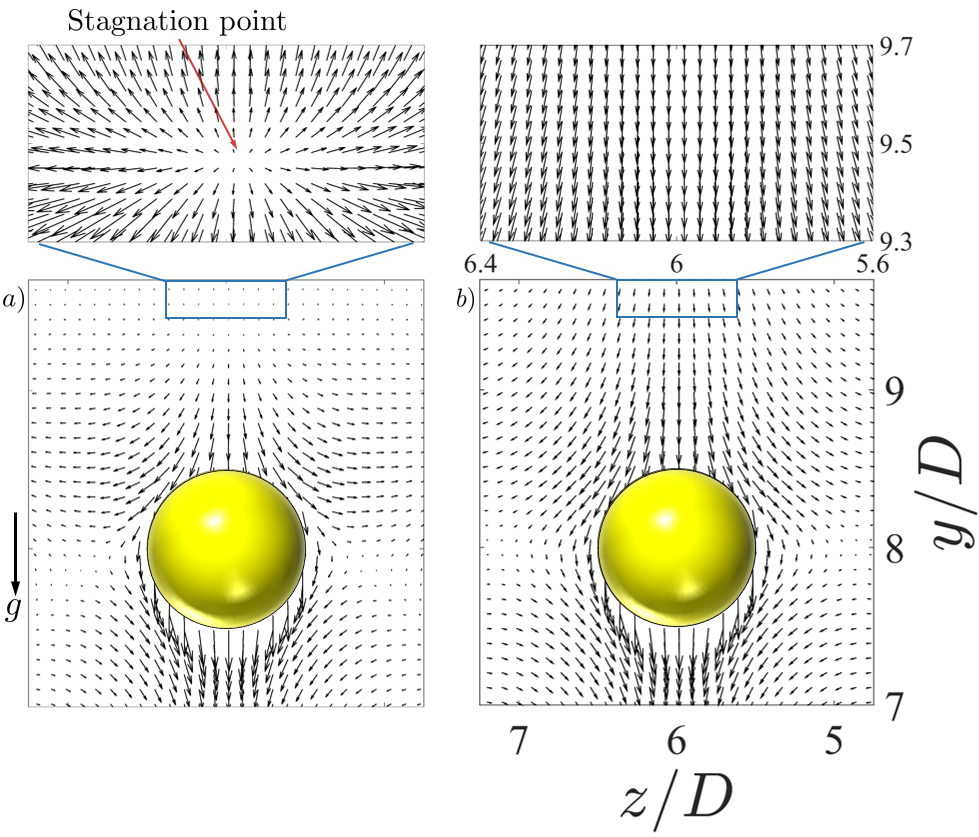} 	
	\caption{Velocity vectors around the sphere settling through an EVP material ($Bi=0.5$ and \rev{$Wi_{\infty}=0.1$}) in the center $yz$ plane for the case of $a)$ pure sedimentation, $b)$ shear-induced sedimentation. The blue boxes magnify the flow upstream of the sphere.}
	\label{fig::vector_sed_SIS}
\end{figure}

Furthermore, the negative wake is eliminated by imposing the cross shear flow, as demonstrated in figure \ref{fig::vector_sed_SIS} by the velocity vector field around the settling sphere. 
\subsubsection{Yielded/unyielded regions}

It should be noted that all the regions that remain unyielded in the case of pure sedimentation in figures \ref{fig::fig4} and \ref{fig::fig5}  yield as soon as the fluid is sheared. Further, in all of the shear-induced sedimentation simulations, the material (either pure visco-plastic or EVP) is yielded when it enters the computational domain. In other words, the components of the polymer stress tensor satisfy the von-Mises yielding criterion at the inlet. These components are obtained by solving equations (\ref{mass})-(\ref{evp}) at steady-state for 
the uniform and Couette flow of Bingham and EVP fluid analytically (see section \ref{sec::boundary conditions} and appendix \ref{App:Bingham} for the analytical solution of the stress tensor components in the case of EVP and Bingham fluid respectively).

\subsubsection{Drag coefficients}
The objective here is to study how the sphere drag coefficient is influenced by the non-linear coupling of the primary uniform flow and the secondary linear cross shear flow. The drag force is computed as the volumetric sum of the IB forces from the numerical implementation and, from this, the drag coefficient using (\ref{eqn::drag_coefficient}).
\par 
\begin{figure}
	\centering
	\includegraphics[trim={0cm 0cm 0cm 0cm},width=0.5\linewidth]{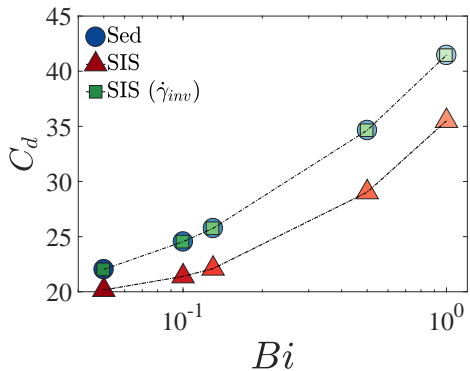} 	
	\vspace{-2ex}
	\caption{Drag coefficient for a settling sphere in an EVP fluid (\rev{$Wi_{\infty}=0.1$}, $Wi=0.01$) for the case of pure sedimentation (blue circular markers) and shear-induced sedimentation (red triangular markers) as a function of the Bingham number. The green square markers represent the drag coefficient deduced from the pure sedimentation simulations data by taking into account the second invariant of the shear rate tensor.}
	\label{fig::fig15}
\end{figure}
\par 
Figure \ref{fig::fig15} shows the drag coefficients for the cases of pure sedimentation (circular markers) and shear-induced sedimentation (triangular markers) of a single sphere in an EVP material. The total drag is an increasing function of the Bingham number, which is similar to the general trend shown in figure \ref{fig::fig8} for the case of pure sedimentation. Clearly, superimposing a cross shear flow on the uniform flow results in the drag reduction on the sphere settling in an EVP fluid (see red triangular markers in figure \ref{fig::fig15}). This is a result of the fact that shearing the yield stress fluid in the orthogonal direction with respect to the uniform flow caused the whole medium to become fully yielded. It is intuitive that the overall viscosity of a yield stress fluid reduces as we impose a cross shear flow and the settling particle experiences 
a lower viscous resistance. In addition, a sphere settling in a yield stress fluid without any cross-shear flow experiences a larger confinement as compared to settling in the presence of external shear flow because the unyielded zones in the EVP fluid act as elastic walls and it is well known that increasing the confinement ratio results in drag enhancement in Newtonian  \cite[see][]{faxen1922widerstand} and viscoelastic fluids  \cite[see e.g.][]{lunsmann1993finite,harlen2002negative}.
\par
\begin{figure}
	\centering
	\includegraphics[trim={0cm 0cm 0cm 0cm},width=0.55\linewidth]{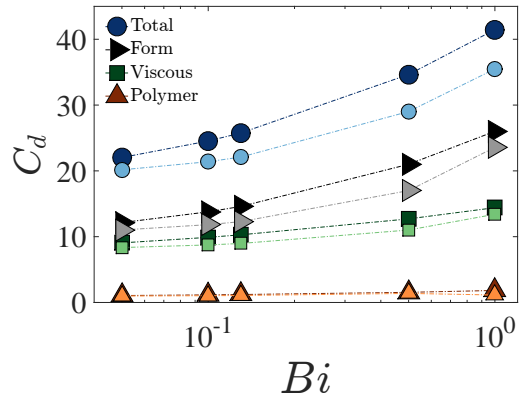} 	
	\vspace{-2ex}
	\caption{Total and individual drag components on the sphere settling in an EVP fluid at \rev{$Wi_{\infty}=0.1$} in the absence of cross shear flow (dark markers) and in the presence of cross shear flow (light markers) as a function of $Bi$ number}
	\label{fig::fig16}
\end{figure}

 
The total and individual drag contributions for the case of shear-induced sedimentation, computed using equations (\ref{eqn:form drag})-(\ref{eqn:polymer drag}), are compared with the results pertaining pure sedimentation in figure \ref{fig::fig16}. 
According to the data in the figure, the form drag is the dominant component also in the case of shear-induced sedimentation, as documented  for a particle settling in the absence of an externally imposed shear flow in figure \ref{fig::fig8}. The viscous drag is the second largest 
contribution \rev{whereas} the polymer drag provides the smallest contribution to the total drag in both cases. 
The form and viscous drag components are both an increasing function of $Bi$  at constant sedimentation Weissenberg number, \rev{$Wi_{\infty}$}, in which
case the polymer drag remains nearly constant as $Bi$ is increased. 

In summary, in pure sedimentation flows, adding small elasticity to the ideal yield stress fluid indirectly modifies the dynamic pressure and viscous stress distributions on the surface of the sphere such that form and viscous drag contributions are responsible for the observed total drag enhancement. Moreover, by superimposing a relatively weak cross-shear flow to the primary settling flow, the form and viscous drag components drop, while the polymer drag remains almost the same.
\par
\begin{figure}
	\centering
	\includegraphics[trim={0.6cm 0cm 0cm 0cm},width=1\linewidth]{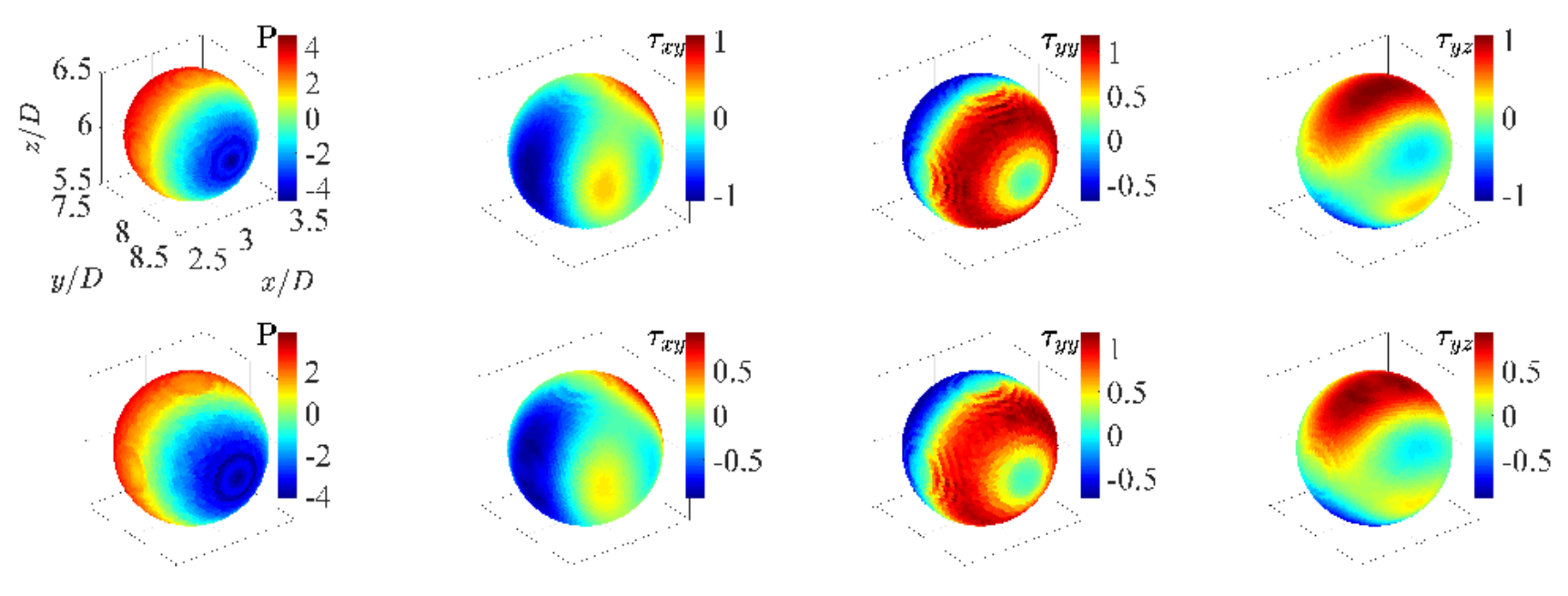} 	
	\vspace{-2ex}
	\caption{Contour plots of the pressure and viscous stress components on the surface of the sphere settling in an EVP fluid at $Bi=1$, \rev{$Wi_{\infty}=0.1$}. The first row corresponds to pure sedimentation ($Wi=0$) and the second row to shear-induced sedimentation ($Wi=0.01$).}
	\label{fig::fig17}
\end{figure}

 
To further clarify the source of drag reduction resulting from the external cross-shear flow, we examine in detail  the case $Bi=1$. 
The distribution of dynamic pressure as well as the viscous stress components that cause drag are therefore displayed in figure \ref{fig::fig17} for pure sedimentation and shear-induced sedimentation. 

%
%
This figure shows that the magnitude of the pressure and viscous shear stresses decrease as soon as the cross-shear flow is superimposed on the uniform flow, whereas, the viscous normal stress ($\tau_{yy}$) remains almost unaffected. 

\subsection{Shear-induced sedimentation in a Bingham fluid}
In this section, we study the effect of superimposing the simple cross shear flow on the orthogonal uniform flow of a Bingham fluid over a single sphere. As the material is sheared, the whole medium becomes yielded and there is no rigid zone left in the medium. 
\par
\begin{figure}
	\centering
	\includegraphics[trim={0.6cm 0cm 0cm 0cm},width=1\linewidth]{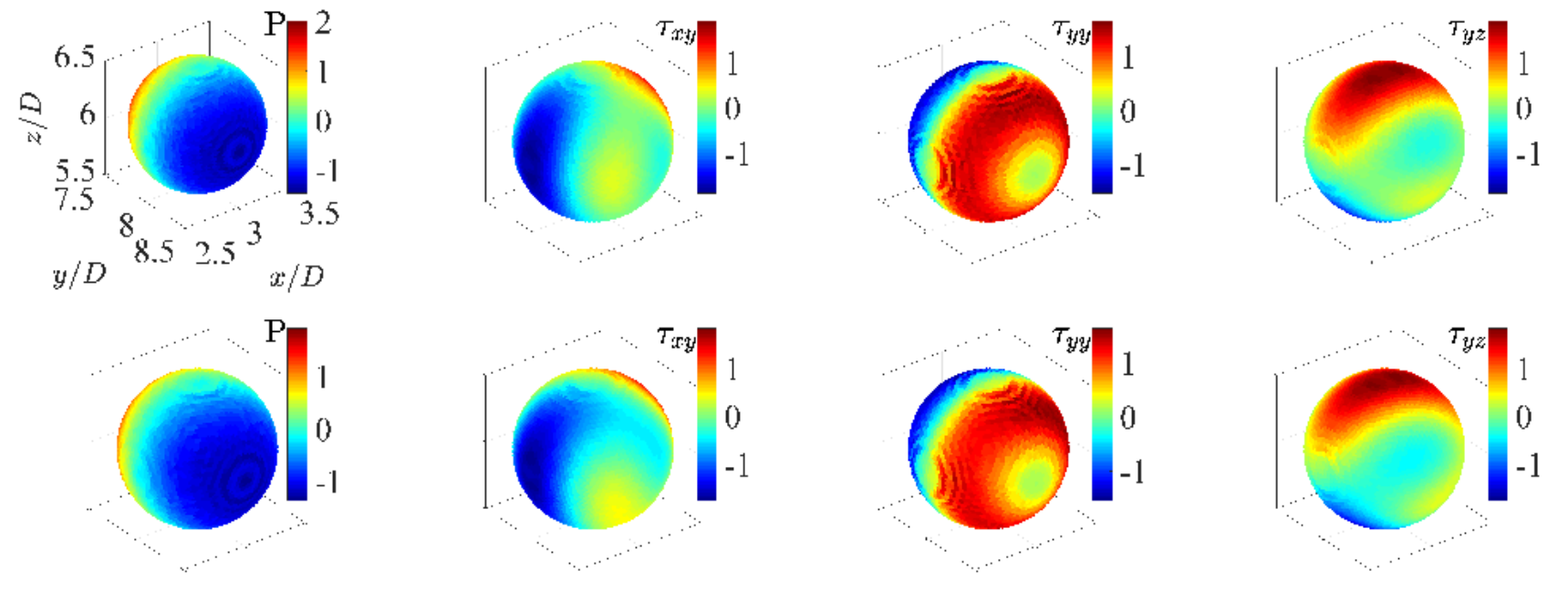} 	
	\vspace{-2ex}
	\caption{Contour plots of the dynamic pressure and of the three components of the viscous stress on the surface of the particle in the case of pure sedimentation (first row) and shear-induced sedimentation (second row) in a Bingham fluid at $Bi=0.05$.}
	\label{fig::fig26}
\end{figure}

 
Table \ref{table::VP_Sed_SIS_Drag} shows the total drag coefficient and its components (form and viscous drag) on the sphere settling in a Bingham fluid in the absence and the presence of an externally imposed cross shear flow.
\begin{table}
	\begin{center}
		\def~{\hphantom{0}}
		\begin{tabular}{l c c}
			Drag & Pure Sedimentation &Shear-induced Sedimentation\\ 
			Total   & 27.15 & 25.55 \\
			Viscous & 15.89 & 14.95 \\
			Form    & 11.26 & 10.60 
		\end{tabular}
		\caption{Comparison of the drag coefficient and its individual components on a sphere settling in the absence and the presence of cross shear flow in a Bingham fluid at $Bi=0.05$.} \label{table::VP_Sed_SIS_Drag}
	\end{center}
\end{table}
In both cases, the viscous drag component is the dominant one. Furthermore, the presence of the cross-shear flow results in the reduction of both viscous and form drag. This behavior can be inferred from the pressure and viscous stress distributions in figure \ref{fig::fig26}.

%
%

\par 
We next compare the drag on a sphere settling in a Bingham fluid to that of a sphere settling in an EVP fluid where both fluids are subjected to cross shear flow at the same $Bi$ number. This would be useful to study the effect of elasticity on the sphere drag settling in sheared yield stress fluid as the elasticity is absent for the Bingham fluid. Table \ref{table::VP_EVP_SIS_Drag} provides this comparison.

\begin{table}
	\begin{center}
		\def~{\hphantom{0}}
		\begin{tabular}{l c c}
			Drag & EVP fluid & Bingham fluid\\ 
			Total   & 20.16 & 25.55 \\
			Form    & 11.02 & 10.60 \\
			Viscous & 8.35 &  14.95 \\
			Polymer & 0.79 &   - \\
		\end{tabular}
		\caption{Comparison of the drag coefficient and its individual components on a sphere settling in sheared visco-plastic (Bingham) fluid at $Bi=0.05$ and EVP material at \rev{$Wi_{\infty}=0.1$}, $Bi=0.05$.} \label{table::VP_EVP_SIS_Drag}
	\end{center}
\end{table}

By introducing the elasticity, the total drag coefficient is reduced. The drag reduction is mainly due to a decrease in viscous drag, while the form drag is hardly affected by the elasticity. Adding polymers to a Bingham fluid indirectly decreases the sphere drag by reducing the viscous stresses on the sphere surface. 

\subsection{Nonlinear Coupling}
The Bingham number can be defined in two ways depending on how one selects the scale of the shear rate that can be either the one induced by the settling of the sphere $(\dot{\gamma}_{sett})$ or by the scalar shear rate that includes both the shear and the settling flow. For the latter choice,  the following second invariant of the deformation rate tensor $(\dot{\gamma}_{inv})$ can be defined. 
\begin{equation} \label{eqn: gamma dot inv}
\dot{\gamma}_{inv}=\sqrt{\dot{\gamma}_{sett}^2+ \dot{\gamma}_{0}^2}=\frac{U_{\infty}}{D}\sqrt{1+\alpha^{2}}.
\end{equation}
This is due to the fact that the cross-shear flow results in an additional off-diagonal element of the strain rate tensor. $Bi$ and $Bi_{inv}$ therefore can be defined by the following relations:
\begin{equation}\label{eqn:Bi sett inv}
Bi=\frac{\tau_{0}}{\eta_{0} \dot{\gamma}_{sett}}; \qquad  Bi_{inv}=\frac{\tau_{0}}{\eta_{0} \dot{\gamma}_{inv}},
\end{equation}
In the shear-induced sedimentation simulations, the shear flow is secondary to the sedimentation flow, and one might try to deduce the drag coefficient from the pure sedimentation simulations data. This means that, having the  relationship between $C_d$  and $Bi$ for settling flows, one can estimate the drag coefficient for shear induced sedimentation flows by replacing $Bi$ with $Bi_{inv}$, with the green square markers in figure  \ref{fig::fig15} represent the result of such an estimation. Because in our simulations, the shear flow is secondary, and consequently,  $Bi\sim Bi_{inv}$ the mentioned estimate results in a drag coefficient that is almost the same for the settling and shear induced sedimentation flows (as in figure  \ref{fig::fig15} the blue circles are very close to the green squares). However, our simulations show that the drag coefficients for the shear induced sedimentation (red triangles in figure  \ref{fig::fig15}) are much smaller than the values of $C_d$ in the pure sedimentation simulations (blue circles in figure  \ref{fig::fig15}). This implies that the coupling of  the two orthogonal flows play a significant role in determining the sphere drag and the mentioned estimate is not valid even when one of the flows is one order of magnitude smaller than the other.
\par 
To expand on the above discussion, our simulations show that the changes in pressure and viscous stresses are not linear with respect to changes in the second invariant of shear rate. For instance at $Bi=0.5$ for a settling flow, a relative increase of $0.5\%$ in  $\dot{\gamma}_{inv}$ by superimposing a secondary shear flow, contributes to approximately a $28\%$ relative decrease in the dynamic pressure and $18\%$ in all of the viscous stress components. This is due to the nonlinear coupling of the secondary simple cross shear flow to the primary uniform flow of the EVP fluid.

\subsection{Shear-induced sedimentation at higher elasticity}
In this section, we show the results of the simulations of a single spherical particle settling in a sheared EVP fluid at higher elasticity, i.e, at \rev{$Wi_{\infty}=1$}. These simulations were performed for the same range of $Bi$ number, i.e, $0.05\leq Bi \leq 1$. Note again that  the ratio between shear and settling Weissenberg number ($\alpha=Wi/\rev{Wi_{\infty}}$) is kept fixed to $\alpha=0.1$. Therefore, the shear Weissenberg number $Wi=0.1$.
\begin{figure}
	\centering
	\includegraphics[trim={0cm 0cm 0cm 0cm},width=0.55\linewidth]{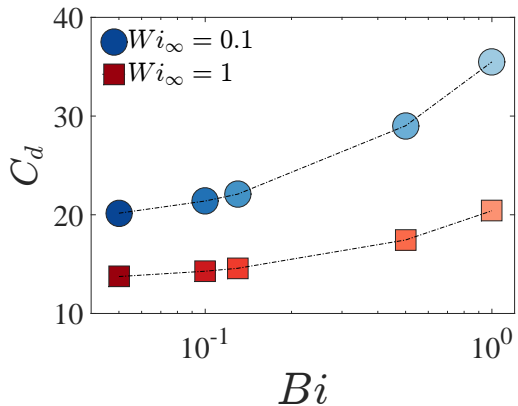} 	
	\vspace{-2ex}
	\caption{Drag coefficient versus the Bingham number for a sphere settling in sheared EVP fluid at different settling Weissenberg numbers, \rev{$Wi_{\infty}$}.}
	\label{fig::fig20}
\end{figure}
\par 
The drag coefficient for a sphere settling in an EVP fluid subject to an externally imposed shear flow at these two different settling Weissenberg numbers is reported in figure \ref{fig::fig20} as a function of $Bi$. For both values of \rev{$Wi_{\infty}$} considered, the drag is an increasing function of $Bi$, while it is a decreasing function of \rev{$Wi_{\infty}$} for all $Bi$ examined. In other words, the drag reduces as the material elasticity increases  at constant plasticity. 
\par 
Drag reduction for a particle settling in an EVP material at higher elasticity was previously observed in the computations of \cite{fraggedakis2016yielding} of pure sedimentation. In this case, the particle translates in a yielded envelope; as the material elasticity is increased, the volume of this envelope increases while the rigid polar caps shrink because the von-Mises yielding criterion is satisfied more easily due to the larger elastic stresses in the medium. Consequently, the elastic walls move further away from the particle surface at larger value of \rev{$Wi_{\infty}$} and the particle experiences lower confinement  at higher elasticity, which results in drag reduction \citep{fraggedakis2016yielding}.
\par 

As far as the mechanism of the drag reduction due to the elasticity in a shear-induced sedimentation case is concerned, we display the total drag as well as its components for \rev{$Wi_{\infty}=0.1,1$} at $Bi=1$ in table \ref{table::EVP_SIS_theta_01_1_Drag}. As the material elasticity is increased, all of the drag components decrease. 
Nevertheless, our simulations show that  the form and viscous drag reduction is more pronounced and they are the dominant drag components.

\begin{table}
	\begin{center}
		\def~{\hphantom{0}}
		\begin{tabular}{lcc}
			Drag    & EVP fluid (\rev{$Wi_{\infty}=0.1$}) & EVP fluid (\rev{$Wi_{\infty}=1$})\\ [6pt]
			Total   & 35.49 & 20.40 \\
			Form    & 23.58 & 11.14 \\
			Viscous & 11.41 & 8.87 \\
			Polymer & 0.50 &  0.39 \\
		\end{tabular}
		\caption{Comparison of the drag coefficient and its individual components on a sphere settling in a sheared EVP fluid at $Bi=1$ for two values of \rev{$Wi_{\infty}$}.} \label{table::EVP_SIS_theta_01_1_Drag}
	\end{center}
\end{table}

\begin{figure}
	\centering
	\includegraphics[trim={0.6cm 0cm 0cm 0cm},width=1\linewidth]{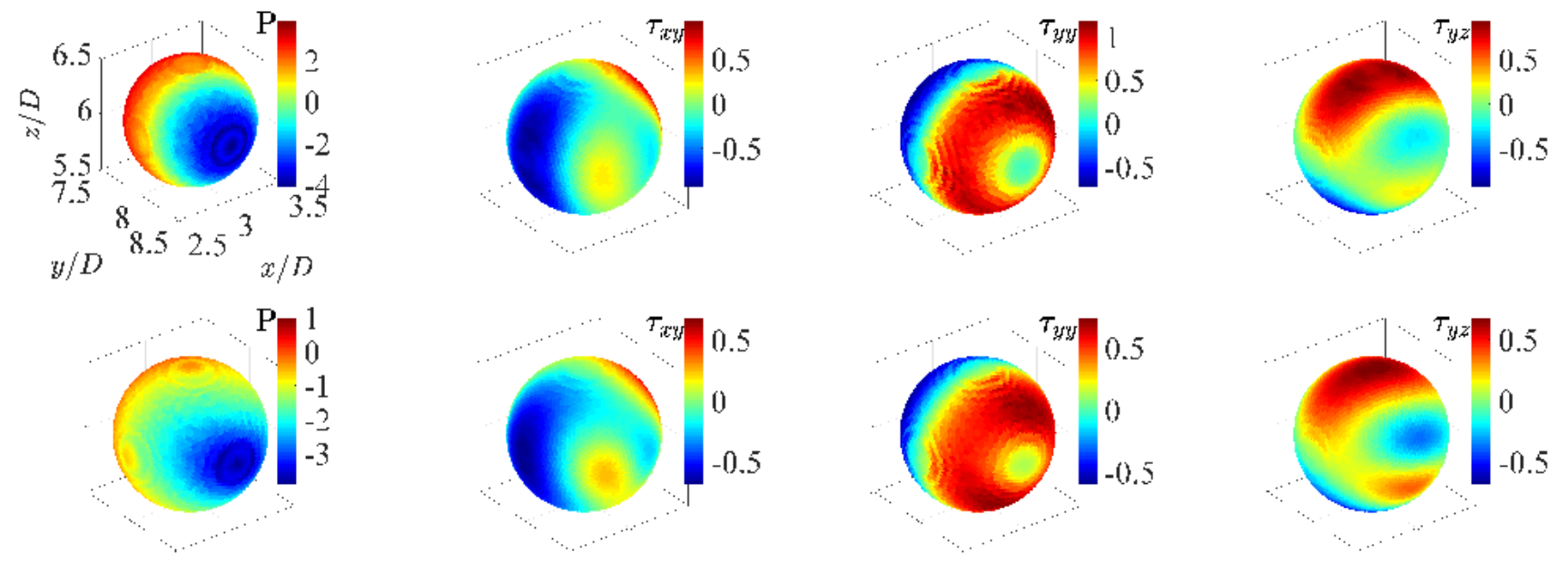} 	
	\vspace{-2ex}
	\caption{Contour plots of pressure and viscous stress components projected on the surface of the sphere settling in sheared EVP fluid at $Bi=1$. The first row corresponds to \rev{$Wi_{\infty}=0.1$} ($Wi=0.01$) and the second row to \rev{$Wi_{\infty}=1$} ($Wi=0.1$).}
	\label{fig::fig24}
\end{figure}

%
%
\par 
To gain further insight, the pressure and viscous stress distributions on the sphere surface are displayed in figure  \ref{fig::fig24} for the same two cases above, i.e.\  for $Bi=1$ and \rev{$Wi_{\infty}=0.1,1$}.
It can be seen that the asymmetry in the positive and negative pressure distributions around the sphere stagnation points at zero polar angle is enhanced at higher elasticity. Moreover, the absolute value of the pressure on the sphere surface is reduced causing form drag reduction at higher elasticity. The normal viscous stress is relaxed at higher shear Weissenberg number. The loss of symmetry of the positive and negative viscous shear stresses ($\tau_{xy}$ and $\tau_{yz}$) is more noticeable around the north and south poles of the sphere, at zero polar angle, and in the equatorial plane on either side of the sphere at zero azimuthal angle. Note that the breaking of the fore-aft symmetry of the velocity magnitude around the sphere by elasticity in a yield stress fluid was observed previously experimentally  \cite[see e.g.][]{holenberg2012particle,firouznia2018interaction} and computationally \citep{fraggedakis2016yielding}.

\section{Conclusion} \label{sec::conclusion}
Direct numerical simulations are performed to study the sedimentation of a single sphere confined in a quiescent and a sheared yield stress fluid at small particle Reynolds number ($Re_{p}=1$).
The 3D simulations are performed for both ideal yield stress fluids (using Bingham model) and EVP materials as the carrying fluid. The former fluids exhibit viscous and plastic behaviors while EVP materials also display elastic effects.
Consequently, we investigate the impact of elasticity on the flow behavior and dynamics of particle settling. 
In all of the simulations the ratio of the constant externally imposed shear rate ($\dot{\gamma_{0}}$) to the shear rate induced by the particle settling ($\dot{\gamma}_{sett}$), defined as the ratio of particle settling velocity to its diameter $(U_{\infty}/D)$ is fixed and equal to $0.1$. For the simulations of  EVP materials, the ratio of the shear to sedimentation Weissenberg number $(Wi/\rev{Wi_{\infty}})$ is held constant and equal to $0.1$.
\par 
The constitutive equations from \cite{saramito2009new} are implemented to model the EVP material. The non-Newtonian stress tensor is fully coupled with the flow equations that are solved with a fast and highly scalable finite-volume method with FFT-based pressure solver. 
The immersed boundary method (IBM), with a computationally efficient multi-direct forcing scheme, is adopted to represent the rigid spherical particle. The no-slip/no-penetration boundary condition on the surface of the particle are therefore implicitly imposed by adding a virtual body force to the right hand side of the momentum equations.
\par 
As concerns the flow and particle dynamics, the fluid velocity distribution around the particle is symmetric with respect to the sphere equatorial plane for pure sedimentation in a Bingham fluid. However, the fore-aft symmetry breaks and the formation of a negative wake observed as the sphere settles in an otherwise quiescent EVP material. 
Since the thixotropy of the yield stress fluid is not considered in the \cite{saramito2009new} constitutive equation for EVP fluids, elasticity is the primary cause of both the fore-aft asymmetry in the velocity field and the negative wake. This is in line with the recent computations of \cite{fraggedakis2016yielding} and several previous experiments \cite[see e.g.][]{gueslin2006flow,putz2008settling,holenberg2012particle}. Here, we show that, superimposing the secondary cross-shear flow, the fore-aft asymmetry in the velocity field becomes less pronounced. Furthermore, the negative wake generated downstream of the sphere during sedimentation in an EVP material disappears in shear induced sedimentation. The present $3D$ numerical solver enables us to extract the yielded isosurfaces around the sphere. The yielded surface approaches the particle surface from the equatorial plane as $Bi$ increases, which eventually causes the sphere to stop settling.
\par 
We have also examined the total drag on the particle, along with its individual contributions. 
The total drag is calculated from the numerical data as the volumetric sum of the IB forces and the drag components computed by performing the numerical integration of the corresponding stress fields on the surface of the sphere. As previously reported by \cite{fraggedakis2016yielding}, the drag coefficient for the settling in an EVP fluid increases when increasing plasticity and decreases when increasing elasticity. This trend is found to hold also in a sheared EVP fluid.
In addition, the drag decreases considerably once the cross-shear flow is superimposed on the uniform flow for both Bingham and EVP material at constant plasticity and elasticity. The key finding is that the drag coefficient for a sphere settling in a sheared EVP fluid cannot be obtained from the drag coefficient pertaining  a sphere settling in an otherwise quiescent fluid. This implies that the coupling of the cross-shear flow and the uniform flow is non-linear and plays a major role in determining the sphere drag in yield stress fluids.
\par 
In this study, as the particle settles in a sheared yield stress fluid, the second invariant of the deformation rate tensor is increased by only $0.5\%$ as compared to settling in the absence of cross-shear flow. This $0.5\%$ increase in the shear rate induces an approximately $5\%$ decrease in dynamic pressure, $4\%$ reduction of viscous normal stress ($\tau_{yy}$), and $3\%$ and $16\%$ decrease in the $\tau_{xy}$ and $\tau_{yz}$ components of the viscous shear stress. 
Interestingly, however, the change in the dynamic pressure and the viscous stresses on the surface of the sphere settling through an EVP material is more than $18\%$ for the whole range of $Bi$ investigated (except for the change in viscous normal stress ($\tau_{yy}$) at $Bi=1$ which is approximately $2\%$). Therefore, the coupling between the cross-shear flow and the uniform flow past a sphere affects the stress fields and the drag significantly.
\par 
By decomposing the total drag coefficient for a sphere settling in an otherwise quiescent EVP fluid into its components, it becomes evident that the form drag (resulting from the dynamic pressure on the particle surface) is the dominant component and the primary cause of drag enhancement with increasing Bingham number. 
Nevertheless, the viscous drag is the largest component in a Bingham fluid at very small $Bi$ number ($Bi=0.05$). The dynamic pressure and stress fields on the surface of the sphere are 
comprehensively analyzed. We find that, for the EVP fluid, as $Bi$ increases, the magnitude of the dynamic pressure also increases while its symmetry around the sphere stagnation points breaks,  resulting in further drag enhancement with the material plasticity. Adding a small degree of elasticity to the purely Bingham fluid modifies the viscous stresses on the surface of the sphere settling in both quiescent and sheared EVP fluids. However, the pressure distribution remains almost unaffected. The viscous stresses are modified so as to give viscous drag reduction. Consequently, the total drag is less in the case of the EVP fluid than for the Bingham fluid, regardless of the existence of an externally imposed cross-shear flow. Thus, adding polymers to an ideal yield stress fluid causes drag reduction through the modification of the viscous stresses on the surface of the particle settling in both quiescent and sheared EVP fluids. 
\par 
We have also performed simulations of a sphere settling in a sheared EVP material at higher elasticity (\rev{$Wi_{\infty}=1$}) and found that the drag coefficient is lower than 
that at lower elasticity (\rev{$Wi_{\infty}=0.1$}). This drag reduction is mainly due to a decrease of both the form and viscous drag components, while the polymer drag remains almost unaffected. 
Moreover, the longer relaxation time of the macromolecular chains affects the normal viscous stresses, reducing their magnitude on the sphere surface. Conversely, the magnitude of the viscous shear stresses ($\tau_{xy},\tau_{yz}$) slightly increases at the particle surface; this is however not sufficient to overcome the decrease in viscous normal stress ($\tau_{yy}$), so that the viscous drag is lower at higher $Wi$. Note also that
 the asymmetry in the dynamic pressure and viscous stress distributions is more pronounced at higher elasticity (\rev{$Wi_{\infty}=1$}).
\par
This study opens an avenue in answering many fundamental questions involving particles in practical yield stress fluids, i.e., the effects of shear thinning and solvent to polymer viscosity ratio on the drag, the drag laws when multi-body interactions between particles are present, confinement effects, lubrication forces between particles, lift forces and particle migration when inertia becomes more relevant, etc. In addition, performing experiments and comparing the results with the simulations, simulations would help us to refine the constitutive laws of practical yield stress fluids as in reality thixotropy, elasticity and plasticity coexist in such materials.  We hope this work will provide new insights to help tackling these challenging problems.

\section*{Acknowledgments}
This work was supported by the National Science Foundation (grant CBET-1554044-CAREER), National Science Foundation's Engineering Research Centers (grant CBET-1641152 Supplementary CAREER), American Chemical Society Petroleum Research Fund (grant 55661-DNI9). The authors acknowledge the computer time provided by the Ohio Supercomputer Center (OSC). L.B. and M.E.R. acknowledge the support from the European Research Council under grant ERC-2013-CoG-616186, TRITOS.

\appendix
\section{Bingham fluid modeling} \label{App:Bingham}
To model the stress-deformation behavior of a visco-plastic fluid, the Bingham constitutive equation has first been proposed \citep{bingham1922fluidity,bird1983rheology}. 
These relations are recovered by setting $Wi=0$, $\beta=1$ and $n=1$ in equations (\ref{mass})-(\ref{evp}).
\par 
In our simulations, all boundary conditions remain the same as described in section (\ref{sec::boundary conditions}) for the case of an EVP material, except the inlet condition where we use the components of the visco-plastic stress tensor. As mentioned in the text, these are obtained by analytically solving the Bingham fluid 
for the combined Couette and uniform flow at steady state in the absence of the particle. In this case, the only non-vanishing component of the stress tensor is the shear stress in the spanwise direction of the shear plane ($\tau_{xz}$) which is obtained via:
\begin{equation}\label{APP:eqn_Bingham_analytical_solution}
\tau_{xz}=2 \alpha +Bi.
\end{equation}
Note that contrary to an EVP material, the first normal stress difference is zero for the case of a Bingham fluid. Here, 
we perform the simulations at two $Bi$ numbers ($Bi=0.05,0.5$) for pure sedimentation and at one $Bi$ number ($Bi=0.05$) for shear-induced sedimentation of a single sphere, with $\alpha=0$ and $0.1$ respectively.

\subsection{Numerical Method}\label{App:Bingham_Numerical}
The Bingham fluid constitutive equation has an inherent discontinuity as the state of stress is undetermined before the material yielding point. To overcome the difficulty associated with the numerical treatment of the yield stress constitutive equation, we use the regularization method with the modification proposed by \cite{papanastasiou1987flows} where 
the exponential growth of the extra stress tensor is controlled by the material parameter, $m$; consequently, we can apply the same equation to the rigid and deformed regions of the medium. 
The viscous stress and the deformation rate tensors are related to each other via the apparent viscosity, 
\begin{equation}\label{eqn:regularization}
\eta_{app}=\eta_{p}+\frac{\tau_{0}}{|\dot{\gamma}|}(1-\exp(-m|\dot{\gamma}|)),
\end{equation} 
where $|\dot{\gamma}|$ is the second invariant of the shear rate tensor and $m$ is the stress growth exponent. The stress tensor is then $\boldsymbol{\tau}=\eta_{app} \boldsymbol{\dot{\gamma}}$.

In order for this equation to mimic the ideal Bingham fluid, $m$ should be chosen to be sufficiently large \cite[see e.g.][]{blackery1997creeping,zisis2002viscoplastic,mitsoulis2004creeping,tokpavi2008very}. \rev{It is worth mentioning that the yield surface location as predicted by the regularized viscosity function is a strong function of the regularization parameter $m$ and the behavior of the true yield-stress fluid is recovered when $\left(m \rightarrow \infty \right)$. 
Here, we have adopted a value of $1000$ for the regularization parameter $m$ except for the validation case when we chose a value of $200$
for consistency with the simulations performed by \cite{blackery1997creeping}.  Moreover, we have checked that results for values of $m=500$  and $m=1000$ are in close agreement for the cases considered.} The stress tensor components are advanced in time with the same Crank-Nicolson scheme used for the EVP fluid and the spatial derivatives computed using second-order central differences. 

\subsection{Validation case} 
\label{App::Validation_Bingham}
We validate our implementation against the results by \cite{blackery1997creeping} for the case of 
the uniform flow of Bingham fluid past a single sphere held stationary in a rectangular channel. 
The computational domain  has a confinement ratio similar to \cite{blackery1997creeping} with domain size
$L_{x}=6D$, $L_{y}=16D$ and $L_{z}=4D$, discretized with $192 \times 512 \times 128$ points. Note that \cite{blackery1997creeping} performed the simulations in a tube by adopting the axisymmetric boundary condition rather than a rectangular channel. The boundary conditions are described in section \ref{sec::boundary conditions}. In our simulation, the particle Reynolds number is chosen to be $1$ and the Bingham number $Bi=0.108$.   
\par 

 \cite{blackery1997creeping} proposed a correlation for the drag of a sphere settling in a Bingham fluid as a function of Bingham number and confinement ratio. Defining the  Stokes drag as:
\begin{equation}\label{eqn:Stokes_drag2}
C_{s}=\frac{2F_{d}}{6 \pi \eta_{p} U_{\infty} D },
\end{equation}
the following correlation is given by \cite{blackery1997creeping}:
\begin{equation}\label{eqn:Stokes_drag_mitsoulis2}
C_{s}=C_{s,N}+aBi^{b},
\end{equation}
where $C_{s,N}$ is the Stokes drag coefficient for a sphere settling in a Newtonian fluid at the same confinement ratio, which can be obtained using Bohlin's approximation \cite[see e.g. ][]{miyamura1981experimental,zheng1991flow}. 
The coefficients  $a$ and $b$ depend on the tube to sphere radius ratio. For a confinement ratio of $4$, these constants take the value of $1.92$ and $0.92$ respectively. Table \ref{table::Mitsoulis} compares the Stokes drag coefficient resulting from the present work with the predictions of equation \ref{eqn:Stokes_drag_mitsoulis2}, showing that
the present results predict the Stokes drag coefficient with a relative difference of less than $0.5\%$. 
\begin{table}
	\begin{center}
		\def~{\hphantom{0}}
		\begin{tabular}{l c c c}
			$L_{z}/D$ & $C_{s}$ (PW) & $C_{s}$ (\cite{blackery1997creeping}) & $\%$ difference \\[3pt]
			4 & 2.235 & 2.227 & 0.360 \\ 
		\end{tabular}
		\caption{Comparison of Stokes drag coefficient of a sphere settling in a Bingham fluid resulting from the present work (PW) and the computations of \cite{blackery1997creeping}.} \label{table::Mitsoulis}
	\end{center}
\end{table}
\par
\begin{figure}
	\centering
	\includegraphics[trim={0cm 0cm 0cm 0cm},width=0.7\linewidth]{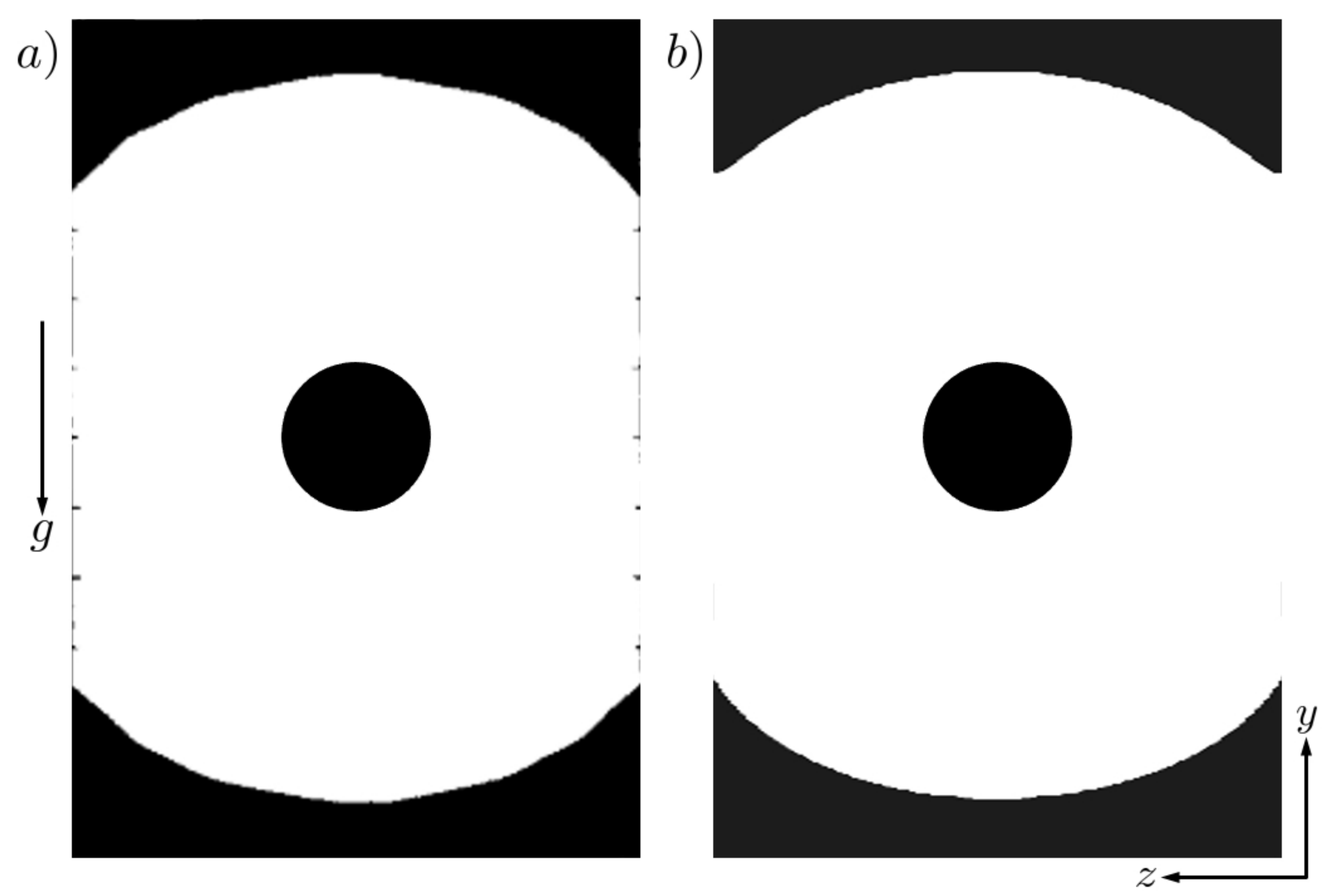} 	
	\caption{Comparison of the yielded/unyielded zones around a particle settling in a Bingham fluid at the confinement ratio of $4$ and at $Bi=0.108$; $a)$ computations of \cite{blackery1997creeping} at $Re_{p}=0$, $b)$ present work at $Re_{p}=1$. White and black colors represent the yielded and unyielded zones respectively.}
	\label{fig::Validation}
\end{figure}

 
In figure \ref{fig::Validation}, 
we compare the size and shape of the yielded/unyielded zones around the sphere with the data from \cite{blackery1997creeping}. 
The agreement between our results at $Re_{p}=1$ and the computations of \cite{blackery1997creeping} at $Re_{p}=0$ is quite satisfactory. The slight zig-zagging of the yielded surface boundary results from adopting a large value for the regularization parameter ($m=200$) in the Bingham fluid constitutive equation, which generates non-smooth lines in the solution. For more details, the reader is referred to \cite{zisis2002viscoplastic}. As the Stokes drag coefficient and the size and shape of the yielded/unyielded zones at $Re_{p}=1$ are in very good  agreement with the results for creeping flow conditions, we expect the effect of inertia to be negligible at $Re_{p}=1$. See section \ref{sec:drag_pure_sedimentation} for further discussion.

\section{Numerical integration procedure}\label{App: A}
In this section, the surface integration procedure implemented to evaluate the integrals in euqations (\ref{eqn:form drag}-\ref{eqn:polymer drag}) along with the interpolation scheme used to compute the pressure and stress fields on the surface of the particle are described in details. 
\par 
In the IB method, the first prediction velocity is interpolated from an Eulerian grid cell (used for the fluid phase) to the Lagrangian grid (used for the particle phase). 
Then, the IBM force, computed on the Lagrangian grid from the difference between the actual particle velocity of each Lagrangian force point and the interpolated first prediction velocity, is interpolated back to the Eulerian grid. These are called \textit{interpolation} and \textit{spreading} operations
\citep{peskin2002immersed,uhlmann2005immersed,breugem2012second} and are performed using the regularized Dirac delta function $(\boldsymbol{\delta_{d}})$ from \cite{roma1999adaptive}. This function, which extends over three grid cells in each coordinate direction, is approximated in three-dimensional space by the following product:
\begin{equation}\label{eqn:dirac3}
\boldsymbol{\delta^{3}_{d}} (\boldsymbol{x}_{ijk}- \boldsymbol{X}_{l})=\delta^{1}_{d}(x_{ijk}-X_{l}) \delta^{1}_{d}(y_{ijk}-Y_{l}) \delta^{1}_{d}(z_{ijk}-Z_{l}),
\end{equation}
where $\boldsymbol{x}_{ijk}$ and $\boldsymbol{X}_{l}$ denote the Eulerian grid point with index $(i,j,k)$ and the Lagrangian marker point position with index $l$. $\delta^{1}_{d}(x_{ijk}-X_{l})$ is an one-dimensional approximation of the delta function: 
\begin{equation}\label{eqn:dirac1}
\delta^{1}_{d}(x_{ijk}-X_{l})=\frac{1}{\Delta x} \phi(\frac{ x_{ijk}-X_{l} }{\Delta x}).
\end{equation}
In equation \ref{eqn:dirac1}, $\Delta x$ is the Eulerian grid size  and $\phi$ is a continuous function that is chosen to satisfy the discrete version of the Dirac delta function properties and is obtained from the following equation \citep{roma1999adaptive}:
\begin{equation} \label{eqn: dirac_phi_function}
\phi(r)=\left\{
\begin{array}{@{}lll@{}}
\vspace{2ex}
\frac{1}{6} (5-3|r|-\sqrt{-3(1-|r|)^{2}+1}) , &  0.5\leq |r| \leq 1.5 \\
\vspace{2ex}
\frac{1}{3}(1+\sqrt{-3 r^{2} +1})   , &  |r|\leq 0.5 \\
0, & otherwise
\end{array}\right.
\end{equation} 
with $r=\frac{ x_{ijk}-X_{l} }{\Delta x}$. The one-dimensional approximations for the delta function in the $y$ and $z$ directions (i.e, $\delta^{1}_{d}(y_{ijk}-Y_{l})$, $\delta^{1}_{d}(z_{ijk}-Z_{l})$) are defined similarly to equation \ref{eqn:dirac1} by replacing the grid spacing, Eulerian and Lagrangian grid point positions in the corresponding coordinate direction. 
\par
If the dimension of the Eulerian grid cell in each direction is spatially uniform then the regularize Dirac delta function proposed by \cite{roma1999adaptive} ensures that the total hydrodynamic force and torque that the fluid and particles exert onto each other are preserved in the interpolation and spreading operations \citep{uhlmann2003first,breugem2012second}. Thus, in the present work the Eulerian grid is considered to be a Cartesian grid with uniform size in each coordinate direction, i.e, $\Delta{x}=\Delta{y}=\Delta{z}$. 
For improved accuracy, the Lagrangian points should be uniformly distributed all over the surface of the sphere, with a spacing of the order of the Eulerian grid. 
In the present work, $3219$ Lagrangian points are therefore used to match the Eulerian grid resolution.

\par 
Since the Lagrangian cells are evenly distributed over the surface of the sphere, the surface of the sphere is partitioned into regions of equal area in such a way that the unity between domains is null, while the union of them constitutes the entire sphere surface area. We use the quadrature rule to evaluate the surface integral  \cite[see e.g.][]{atkinson1982numerical,reeger2016numerical,an2016numerical}. The center position of each Lagrangian partition is chosen as the quadrature node. Hence, the weight of the corresponding quadrature rule is positive and equal, and the integral of any quantities over the surface of the sphere can be estimated by the following equation:
\begin{equation} \label{eqn:suface_integral}
\int \hspace{-1.5ex} \int_{S} f n_{j}dS \approx \sum_{l=1}^{N_{l}} f_{l} n_{j,l} S_{l},
\end{equation}
where $f_{l}$ can be any quantity (e.g. pressure, viscous and polymer stresses) of the $l^{th}$ force point, $l$ is the Lagrangian point index, $N_{l}$ is the total number of Lagrangian grid points (total number of equally partitioned subareas on the surface), $n_{j,l}$ is the unit normal vector directed outward on the $l^{th}$ Lagrangian point of the sphere. $S_{l}$ is the surface area of the $l^{th}$ partition which is equal to $\frac{4 \pi R^{2}}{N_{l}}$ for a sphere with radius $R$. 
\par 
Therefore, we only need  the pressure, viscous/polymer stresses, and the unit normal vector on each Lagrangian force point to compute the surface integrals in equations \ref{eqn:form drag}-\ref{eqn:polymer drag} by using equation \ref{eqn:suface_integral} to find the drag components. \\

Computing the form, viscous, and polymer drag components  is performed in several steps. In the first step, the solid volume fraction ($\alpha_{ijk}$) in the grid cell with index $(i,j,k)$ around the collocation points of the velocities ($u,v,w$) and  the pressure ($p$) are computed. This should be done at each Runge-Kutta step.
\par
\begin{figure}
	\centering
	\includegraphics[trim={0cm 0cm 0cm 0cm},width=0.6\linewidth]{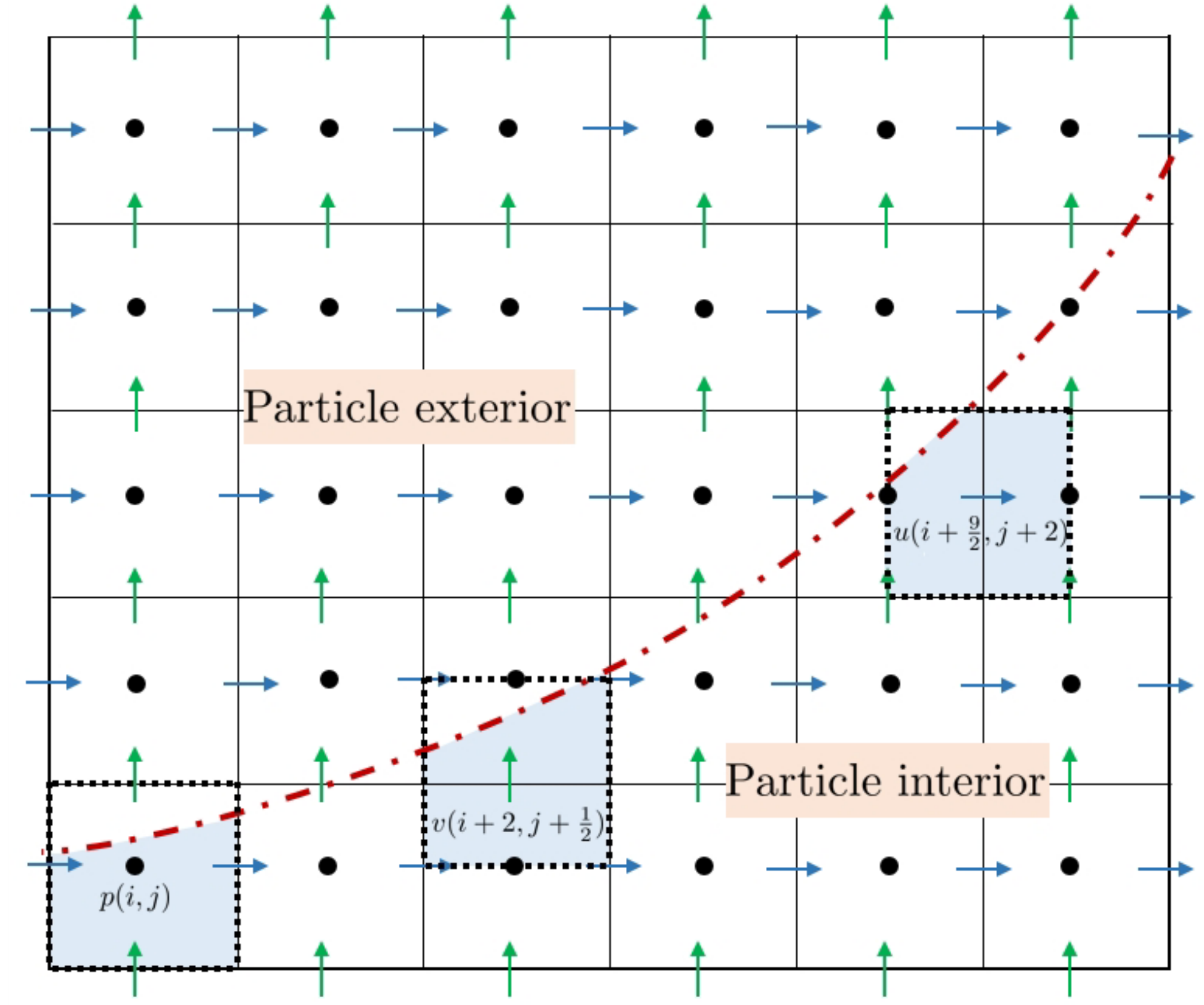} 	
	\caption{Solid volume fractions (highlighted area) for grid cells around $p(i,j)$, $u(i+\frac{9}{2},j+2)$ and $v(i+2,j+\frac{1}{2})$. The sphere boundary is shown by the red dashed line.}
	\label{fig::figA2}
\end{figure}

 
The solid volume fraction $\alpha_{ijk}$ in the Eulerian cubic grid cell with index $(i,j,k)$ is determined from the level-set function $\psi$ obtained by calculating the signed distance of each Eulerian grid point to the particle surface $S$ \citep{kempe2009modelling}: 
\begin{equation}\label{eqn:solid_volume_fraction}
\alpha_{ijk}=\frac{\sum_{k=1}^{8} -\psi_{k} H(-\psi_{k})}{\sum_{k=1}^{8} |\psi_{k}|}.
\end{equation}
In equation \ref{eqn:solid_volume_fraction}, the sum is over all $8$ corner nodes of the cubical Eulerian grid cell and $H$ is the Heaviside step function. The signed distance function $\psi$ represents the position of the Eulerian grid cell with respect to the sphere with $\psi>0$ outside and $\psi<0$ inside the particle. 
Since we use a fully-staggered Cartesian grid, the velocity components are computed at the cell faces while the pressure and the stress fields are calculated at the cell vertex. Hence, the collocation points of the $\zeta$-component of the velocity collocation points in the $\gamma$ direction and the collocation points for the pressure are given by the following equations:
\begin{equation}\label{eqn: velocity_points}
x^{\zeta}_{\gamma}(k)=(k-1-\frac{1}{2}\delta_{\gamma \zeta})\Delta x_{\gamma}; \qquad k=1,2,...,n_{\gamma}+\delta_{\gamma \zeta}.
\end{equation}  
\begin{equation}\label{eqn: pressure_points}
x^{p}_{\gamma}(k)=(k-1) \Delta x_{\gamma}; \qquad k=1,2,...,n_{\gamma}.
\end{equation}
In equations \ref{eqn: velocity_points} and \ref{eqn: pressure_points}, $\Delta x_{\gamma}$ is the size of the grid cell in the $\gamma$ direction and $n_{\gamma}$ is the number of pressure grid points in the $\gamma$ direction.
The use of the staggered grid results in different solid volume fractions depending on the variable considered. Figure \ref{fig::figA2} depicts three different solid volume fractions (highlighted areas) for the grid cells around $p(i,j)$, $u(i+\frac{9}{2},j+2)$ and $v(i+2,j+\frac{1}{2})$. The sphere boundary is shown by the red dashed line.

\par 
Finally, the fluid velocity in each coordinate direction ($u,v,w$) and the pressure at each Runge-Kutta step are obtained by:
\begin{equation}\label{eqn::ushow}
u^{fq}_{ijk}=(1-\alpha^{u}_{ijk})u^{q}_{ijk},
\end{equation}
\begin{equation}\label{eqn::vshow}
v^{fq}_{ijk}=(1-\alpha^{v}_{ijk})v^{q}_{ijk},
\end{equation}
\begin{equation}\label{eqn::wshow}
w^{fq}_{ijk}=(1-\alpha^{w}_{ijk})w^{q}_{ijk},
\end{equation}
\begin{equation}\label{eqn::pshow}
p^{fq}_{ijk}=(1-\alpha^{p}_{ijk})p^{q}_{ijk},
\end{equation}
where $u^{fq}_{ijk}$,$v^{fq}_{ijk}$, $w^{fq}_{ijk}$ and $p^{fq}_{ijk}$ are the velocity components in the $x$, $y$ and $z$ directions and the pressure after direct accounting for the inertia of the fluid contained within the sphere at the Runge-Kutta step $q$. $\alpha^{u}_{ijk}$,$\alpha^{v}_{ijk}$,$\alpha^{w}_{ijk}$ are the solid volume fractions around the velocity collocation points in the $x$, $y$ and $z$ directions (Eulerian cell face) and $\alpha^{p}_{ijk}$ is the solid volume fraction around the pressure collocation point that is at the Eulerian grid cell vertex (see figure \ref{fig::figA2}). $u^{q}_{ijk}$,$v^{q}_{ijk}$, $w^{q}_{ijk}$ and $p^{q}_{ijk}$ are the velocity components in the $x$, $y$ and $z$ directions and the pressure before direct accounting of the inertia of the fluid in the volume of the particle at the $q^{th}$ Runge-Kutta step. It should be noted that, as the polymer stresses are computed at the cell center, the same solid volume fraction that is obtained around the pressure collocation point is used for the polymer stress components.
\par 
In the second step, the viscous stresses are computed on each grid cell
\begin{equation}\label{eqn:viscous_stresses}
\tau^{v}_{ij}=(1-\beta) (\frac{\partial u^{fq}_{i}}{\partial x_{j}}+\frac{\partial u^{fq}_{j}}{\partial x_{i}}); \hspace{4ex} i,j=1,2,3,
\end{equation}
where the spatial derivatives are estimated with the central-differencing scheme.
\par 
So far the pressure, viscous and polymer stresses are computed at each Eulerian grid cell with index $(i,j,k)$ considering the solid volume fraction around the pressure and velocity component collocation points. In the third step, the pressure, viscous and polymer stresses should be projected to the corresponding Lagrangian grid cell. This interpolation is done using the same Dirac delta function defined in equations \ref{eqn:dirac3}-\ref{eqn: dirac_phi_function}
\begin{equation}\label{eqn: press_projection}
P_{l}=\sum_{ijk}^{} p^{fq}_{ijk} \boldsymbol{\delta^{3}_{d}} (\boldsymbol{x}_{ijk}- \boldsymbol{X}_{l}) \Delta x \Delta y \Delta z; \qquad l=1,2,...,N_{l},
\end{equation}

\begin{equation}\label{eqn: viscous_projection}
\tau^{v,l}_{pq}=\sum_{ijk}^{} \tau^{v,ijk}_{pq} \boldsymbol{\delta^{3}_{d}} (\boldsymbol{x}_{ijk}- \boldsymbol{X}_{l}) \Delta x \Delta y \Delta z; \qquad l=1,2,...,N_{l}; \qquad p,q=1,2,3,
\end{equation}

\begin{equation}\label{eqn: polymer_projection}
\tau^{Po,l}_{pq}=\sum_{ijk}^{} \tau^{Po,ijk}_{pq} \boldsymbol{\delta^{3}_{d}} (\boldsymbol{x}_{ijk}- \boldsymbol{X}_{l}) \Delta x \Delta y \Delta z; \qquad l=1,2,...,N_{l}; \qquad p,q=1,2,3,
\end{equation}
where $P_{l}$, $\tau^{v,l}_{pq}$ and $\tau^{Po,l}_{pq}$ are the interpolated pressure, viscous and polymer stress components at the Lagrangian point $l$.
\par 
In the last step, we compute the unit normal vector on the $l^{th}$ Lagrangian point. This is easily done once the  
 position of the center of the particle is known
\begin{equation} \label{eqn:unit_normal_vector}
n_{j,l}=\frac{\boldsymbol {\nabla} G}{|\boldsymbol {\nabla} G|},
\end{equation}
where
\begin{equation} \label{eqn: sphere equation}
G= (X_{l}-X_{c})^{2}+ (Y_{l}-Y_{c})^{2} + (Z_{l}-Z_{c})^{2} -R^{2}; \qquad l=1,2,...,N_{l},
\end{equation}
and 
$X_{c}$, $Y_{c}$ and $Z_{c}$ are the coordinates of the sphere centre.
Finally, we are able to estimate the pressure, viscous and polymer stresses at the Lagrangian points. 
Given the unit normal vector, the form, viscous, and polymer drag components (equations \ref{eqn:form drag}-\ref{eqn:polymer drag}) are found from the following relations:
\begin{equation}\label{eqn:form_drag_integral}
\int \hspace{-1.5ex} \int_{S} p n_{y}dS \approx \frac{4 \pi R^{2}}{N_{l}} \sum_{l=1}^{N_{l}} P_{l} n_{y,l}.
\end{equation}

\begin{equation}\label{eqn:viscous_drag_integral}
(1-\beta)\int \hspace{-1.5ex} \int_{S} (\frac{\partial u_{y}}{\partial x_{q}}+\frac{\partial u_{q}}{\partial y}) n_{q} dS \approx (1-\beta) \frac{4 \pi R^{2}}{N_{l}} \sum_{l=1}^{N_{l}} \tau^{v,l}_{yq} n_{q,l}.
\end{equation}

\begin{equation}\label{eqn:polymer_drag_integral}
\int \hspace{-1.5ex} \int_{S} \tau_{yq}n_{q} dS \approx \frac{4 \pi R^{2}}{N_{l}} \sum_{l=1}^{N_{l}} \tau^{Po,l}_{yq} n_{q,l}.
\end{equation}

\subsection{Validation case}
In order to validate the numerical integration scheme that we have implemented in this study, we compute the drag and its components in the case of the Newtonian fluid past a sphere in a finite domain. 
For this test, the domain size is  $L_{x}=6D$, $L_{y}=8D$ and $L_{z}=5D$, with 
inflow-outflow boundary conditions applied in the streamwise $y$ direction, no-slip enforced in the wall-normal $z$ direction and the periodic boundary conditions  in the spanwise $x$ direction. The simulation is performed at $Re_{p}=1$.
\par
In this simulation the total Stokes drag along with the form and viscous drag components are computed and compared with the theoretical solution \citep{leal2007advanced}.
Confinement effects are taken into account with the wall-correction factor given by \cite{miyamura1981experimental}. Note that at the confinement ratio of $0.2$, the wall-correction given by \cite{miyamura1981experimental} converges to the Faxen law \citep{faxen1922widerstand}. 
A comparison between the present results and the \cite{faxen1922widerstand} approximation is presented in table \ref{table::2}, 
where the Stokes drag coefficient
\begin{equation} \label{eqn:stokes drag}
C_{s}=\frac{F_{d}}{6\pi \eta U_{\infty} R},
\end{equation}
with $\eta$ the viscosity of the Newtonian fluid. 
\begin{table}
	\begin{center}
		\def~{\hphantom{0}}
		\begin{tabular}{lcccccc}
			$D/L_{z}$ & $C_{s}$ (PW) & $C_{s}$ \cite[][]{faxen1922widerstand} & $C_{s}^{f}$ (Theory) & $C_{s}^{v}$ (Theory) & $C_{s}^{f}$ (PW) & $C_{s}^{v}$ (PW)  \\[3pt]
			0.2 & 1.65 & 1.67 & 0.55 & 1.12 & 0.56 & 1.09 \\ 
		\end{tabular}
		\caption{Comparison of drag coefficient and drag components from the present work (PW) with Bohlin's approximation and theoretical solutions for Newtonian fluids.} \label{table::2}
	\end{center}
\end{table}

The agreement in the total Stokes drag is satisfactory with a relative difference of approximately $1.5\%$. The relative difference in the form drag and the viscous drag is around $2\%$ and $3\%$ respectively. The small discrepancy between the numerical integration of the dynamic pressure and the stress fields on the surface of the sphere and the theoretical predictions is primarily due to the Lagrangian grid resolution. 
Increasing the number of Lagrangian force points, the integration scheme becomes more accurate and converges towards the theoretical solution.

\bibliographystyle{jfm}
\bibliography{Ref}

\end{document}